\documentclass[11pt]{article}
\usepackage[left=2cm,top=4cm,right=2cm,bottom=3cm]{geometry}
\usepackage[utf8]{inputenc}
\usepackage{amsmath}
\usepackage{amssymb}
\DeclareMathOperator{\sech}{sech}
\usepackage{graphics,subfigure,epsf,epsfig,color}
\usepackage{slashed}
\usepackage{feynmf}
\usepackage{graphicx}
\usepackage{graphics}
\usepackage{epstopdf}
\usepackage{subfigure}
\usepackage{amsfonts}
\usepackage{sectsty}
\usepackage{hyperref}
\usepackage{cite}
\usepackage{pdflscape}
\usepackage{lipsum} 
\begin{document}
\date{}
\title{Testing the behaviour of exotic matter near wormhole throat in $f(R,T)$ gravity}
\maketitle
\begin{center}
   \author{Payel Sarkar}\footnote{payelsarkar305@gmail.com}
\end{center}

\begin{center}
 Birla Institute of Technology and Science-Pilani, K. K. Birla Goa campus, NH-17B, Zuarinagar, Goa-403726, India
 \end{center}
 \vspace*{0.1in}

\abstract{ In this paper, I propose a static wormhole model within modified $f(R,T)$ gravity where $f(R,T)=R+2\lambda T$. The wormhole solutions have been evolved in four cases- three different shape function along with redshift $\phi=\frac{\phi_0}{r}$ and a variable EoS parameter $\omega(r)$ with constant redshift function. I also have explored the energy conditions and the behaviour of exotic matter within the wormhole in all scenarios. Presence of exotic matter violates necessary energy conditions near wormhole throat which gives constraint on modified gravity parameter $\lambda$ in all different cases.}\\

\section{Introduction}
Wormholes are the special solution of Einstein's field equation that connects two distinct points in space-time \cite{Morris,Visser, Visser2} via a tunnel with a mouth where the tunnel has minimal radius, called throat. This throat can have constant radius known as  static wormhole and wormholes with variable throat radius is called as non-static or cosmological wormholes. Although Wormhole is not observed till now \cite{Lemos, Jordan} but attempts to achieve so were proposed \cite{Tsukamoto,Tsukamoto2,Zhou,Rahaman,Kuhfittig,Bambi,Li,Nandi,Harko}. Wormhole was predicted theoretically a long time ago by Einstein and Rosen \cite{Einstein}. after many years Morris and Throne came up with new ideas of Traversible Wormhole which violates the energy conditions if it is filled with some exotic matter \cite{Morris,Hochberg, Hochberg2}. Dark energy or Phantom energy which has negative pressure,is an alternative of exotic matter as they also can violate the energy conditions \cite{Morris2}. The study of Phantom matter in Wormhole geometry has already been shown in many literatures \cite{Jamil, Lobo, Rahaman2 ,Wang, Lobo3, James}.\\

 The throat is the fundamental concept in Wormhole theory and it should satisfy the flaring out condition. Even throat can be filled with normal matter in presence of modified gravity or some high order curvature terms \cite{Rosa}. Wormholes in absence of exotic matter is studied in modified gravity in many literatures.\cite{Lobo2,Godani,Harko2}.
 Hence testing the wormhole geometry in modified gravity is very important in theoretical physics. A static spherically symmetric wormhole solution has been examined in $f(R)$ Gravity \cite{sharif}, in $f(R,T)$ Gravity\cite{Zubair,Godani2, Godani3, Abbas, Pradhan}, in $f(G,T)$ Gravity etc \cite{Sharif2},  Einstein-Cartan Theory where both the mass and spin of matter fields coupled with geometry of space-time\cite{Grezia}. In Ref.\cite{Battista}, stability of the wormhole sourced from casimir energy density is properly discussed in  Rastall-Rainbow gravity which opens a new insights in the field of wormhole geometries.\\
 $f(R,T)$ gravity is an extension of $f(R)$ gravity which includes the Ricci scalar $R$ along with $T$ trace of energy momentum tensor. This theory has been explored in different areas such as compact objects \cite{Moreas, Zubair3,Shamir}, thermodynamics\cite{Momeni}, cosmology \cite{Moraes4, Moraes5, Correa, Alvarenga, Myrzakulov, sharma, Nagpal, Debnath,Ahmed} etc.  Wormhole solution have been evolved in many literatues in the context of $f(R,T)$ gravity using constant redshift function\cite{Mishra, Sahoo, Sahoo2, Parbati}. An exponential shape function has been studied for wormhole both in $f(R,T)$ Gravity and $f(R)$ Gravity \cite{Samanta, Mishra3, Parsaei, Mishra4}. In \cite{Elizalde}, wormholes of two types of varying Chaplygin gas has been studied in $f(R,T)$ gravity along with their energy conditions.  Stability conditions of charged wormhole in f(R,T) gravity using generalized Chaplygin gas is also studied in \cite{Bhatti1}.
 An exponential $f(R,T)$ model has been considered in \cite{Bhatti} and obtain energy conditions for WH. A changed WH solutions has also been discussed in \cite{Moraes3}.  Wormholes can be constructed from non minimal theory also \cite{Garcia,  Garcia2}. Different type of Wormhole solution have been derived considering different type of energy density functions \cite{Elizalde2}. Wormhole solution also studied by many researchers for different form of $f(R,T)$ functions \cite{Moraes6, Godani3,Moraes7}. In \cite{Mishra3, Mishra4} wormhole solution in $f(R)$ gravity has been discussed for shape function $b(r)=r_0\left(\frac{a^r}{a^{r_0}}\right)$ and $r_0+ar_0\left(\frac{1}{r}-\frac{1}{r_0}\right)$ for constant redshift function whereas in \cite{Rahaman3} have provided wormhole solution for $b(r)=m\tanh{(nr)}$ in Einstein gravity . In this manuscript I have concerned in modelling Wormhole in $f(R,T)=R+2\lambda T$ gravity for the above mentioned shape functions with specific redshift function $\phi(r)=\frac{\phi_0}{r}$. Moreover I also have considered a new form of EoS parameter $\omega(r)=1+\ln{(r)}$ for constant redshift.  \\
 
  The present article is organised as follows: in sec 1. I have described the conditions of shape function followed by the derivation of field equations for wormhole in $f(R,T)$ gravity in section 2. Next I have considered here three different types of shape function-i) $b(r)=m \tanh{(nr)}$ ii) $b(r)= r_0+ar_0\left(\frac{1}{r}-\frac{1}{r_0}\right)$, iii)$b(r)=r_0\left(\frac{a^r}{a^{r_0}}\right)$ for a specific red-shift fuction and in case iv) I have taken a varying EoS parameter $\omega(r)=1+\ln{(r)}$ for constant redshift. Finally I have analyzed all the energy conditions for different cases and conclude my results.

\section{Wormhole metric and its conditions}\label{section2}
The static and spherically symmetric Wormhole (WH) metric is defined as,
\begin{equation}
 ds^2=-e^{-2\phi(r)}dt^2+\frac{dr^2}{1-\frac{b(r)}{r}}+r^2d\theta^2+r^2Sin^2\theta d\phi^2
 \label{metric}
\end{equation}

where $\phi(r)$ and $b(r)$ are the redshift function and  the shape fuction respectively which should obey the following conditions:
\begin{itemize}
 \item The radial co-ordinate $r$ lies within $r_0\leq r<\infty$ where $r_0$ is the minimum throat radius.
 \item At the throat $r=r_0$ the shape function will be,
 \begin{equation}
  b(r_0)=r_0
 \end{equation}
 and outside the throat $r>r_0$, $1-\frac{b(r)}{r}>0$
 
 \item The shape fuction $b(r)$ should obey the Flaring out condition i.e.
 \begin{equation}
  b'(r_0)<1
 \end{equation}
\item For asymptotic flat  spacetime the limit $\frac{b(r)}{r}\rightarrow 0$ as $r\rightarrow \infty$.
\item The redshift function $\phi(r)$ must be non vanishing and finite at the throat.
\end{itemize}

\section{Field equations for Wormholes in $f(R,T)$ gravity}

The Einstein-Hilbert action of the modified $f(R,T)$ theory of gravity, in presence of matter can be written as,
\begin{equation}
    \mathcal{S} = \frac{1}{16 \pi G} \int d^4x \sqrt{-g}~f(R,T)+ \int d^4x\sqrt{-g}~\mathcal{L}_m
    \label{action}
\end{equation}
where $R$ is the Ricci Scalar, $T$ is the trace of energy-momentum tensor, $f(R,T) $ is an arbitrary function of $R$ and $T$, $g$ is the determinant of the metric tensor $g_{\mu\nu}$ and G is the Newtonian constant of Gravitation. \footnote{We have used natural units, $c = \hbar = 1$ and considered $8\pi G=1$}.  $\mathcal{L}_m$ is the matter Lagrangian and is  related to the energy-momentum tensor $T_{\mu\nu}$ as,
\begin{equation}
    T_{\mu\nu}=-\frac{2}{\sqrt{-g}}~\frac{\delta}{\delta g^{\mu\nu}}(\sqrt{-g}\mathcal{L}_m)
\end{equation}
By taking the metric variation of the action (\ref{action}), we find the modified Einstein equation as
\begin{equation}
    f_R(R,T)~R_{\mu\nu}-\frac{1}{2}f(R,T)~g_{\mu\nu}+(g_{\mu\nu}\Box-\nabla_{\mu}\nabla_{\nu})~f_R(R,T)=T_{\mu\nu}-f_T(R,T)~T_{\mu\nu}-f_T(R,T)~\theta_{\mu\nu}
    \label{field}
\end{equation}
Here I have taken $\mathcal{L}_m=-\rho$, hence $\theta_{\mu\nu}=-2T_{\mu\nu}-\rho g_{\mu\nu}$. Considering $f(R,T)=R+2\lambda T$, $\lambda$ is a constant, Eq.~(\ref{field}) will be,
\begin{equation}
 R_{\mu\nu}-\frac{1}{2}Rg_{\mu\nu}=T_{\mu\nu}+\lambda Tg_{\mu\nu}+2\lambda T{\mu\nu}+2\lambda\rho g_{\mu\nu}
 \label{Einstein}
\end{equation}
I have taken an anisotropic fluid for matter content of the form
\begin{equation}
 T{\mu}^{\nu}=diag(-\rho,P_r, P_t,P_t)
 \label{tmunu}
\end{equation}
where $\rho(r)$ is the energy density, $P_r(r)$ , $P_t(r)$ are the radial and lateral pressure respectively. The trace of energy-momentum tensor will be $T=\rho+P_r+2P_t$.\\
The field equations for the metric Eq.(\ref{metric}) along with Eq.(\ref{tmunu}) are,
\begin{equation}
 \frac{b'}{r}=(1+\lambda)\rho-\lambda(P_r+2P_t)
\end{equation}
\begin{equation}
 -\frac{b}{r^3}+\frac{2}{r}\phi'\left(1-\frac{b}{r}\right)=\lambda\rho+P_r(1+3\lambda)+2\lambda P_t
\end{equation}
and 
\begin{equation}
 \left(1-\frac{b}{r}\right)\left(\phi''+\phi'^2+\frac{\phi'}{r}-\phi'\frac{b'r-b}{2r(r-b)}-\frac{b'r-b}{2r^2(r-b)}\right)=\lambda\rho+\lambda P_r+P_t(1+4\lambda)
\end{equation}
From the above three field equations we can get the energy density, radial pressure and lateral pressure as,
\begin{equation}
 \rho=\frac{b'}{r^2(1+2\lambda)}+\frac{2\lambda}{(1+4\lambda)(1+2\lambda)}\left(1-\frac{b}{r}\right)\left(\phi''+\phi'^2-\frac{\phi'}{r}-\phi'\frac{b'r-b}{2r(r-b)}\right)+\frac{6\lambda}{(1+4\lambda)(1+2\lambda)}\frac{\phi'}{r}\left(1-\frac{b'}{r}\right)
 \label{rho}
\end{equation}
\begin{equation}
 P_r=-\frac{b}{r^3(1+2\lambda)}+\frac{1+\lambda}{(1+4\lambda)(1+2\lambda)}\frac{2}{r}\phi'\left(1-\frac{b}{r}\right)-\frac{2\lambda}{(1+2\lambda)(1+4\lambda)}\left(1-\frac{b}{r}\right)\left(\phi''+\phi'^2-\frac{\phi'}{r}-\phi'\frac{b'r-b}{2r(r-b)}\right)
 \label{pr}
\end{equation}
and,
\begin{equation}
 P_t=\frac{b-b'r}{2r^3(1+2\lambda)}+\frac{1}{1+4\lambda}\left(1-\frac{b}{r}\right)\left(\phi''+\phi'^2-\frac{\phi'}{r}-\phi'\frac{b'r-b}{2r(r-b)}\right)+\frac{1+\lambda}{(1+4\lambda)(1+2\lambda)}\frac{2}{r}\phi'\left(1-\frac{b}{r}\right)
 \label{pt}
\end{equation}

For constant redshift function $\phi=\phi_0$ the above three equation reduces to,
\begin{equation}
 \rho=\frac{b'}{r^2(1+2\lambda)}, ~~P_r=-\frac{b}{r^3(1+2\lambda)}, ~~P_t=\frac{b-b'r}{2r^3(1+2\lambda)}
 \label{redshift}
\end{equation}
\section{Energy conditions for different models}
Energy conditions are important in order to study the behaviour of matter inside the wormhole. The four fundamental energy conditions in GR are as follows-
\begin{itemize}
    \item Null energy condition(NEC) satisfies $\rho+P_i\geq 0$ and it represents the attractive nature of gravity.
    \item Weak energy condition (WEC) satisfies $\rho+P_i\geq 0$, $\rho\geq 0$ and it additionally ensures the positive behaviour of energy density along with NEC.
    \item Strong energy condition (SEC) satisfies $\rho+P_i\geq 0, \rho+\Sigma_iP_i\geq 0$. It originates from attractive nature of gravity and is a result of spherically symmetrical metric. 
    \item Dominant energy condition (DEC) which is $\rho\geq |P_i|$, measures the velocity of energy transfer to the speed of light.
\end{itemize}
In this section, I have discussed the above mentioned energy conditions for four different cases of matter present at the wormhole.
\subsection{Case 1: $\phi(r)=\frac{\phi_0}{r}$, $b(r)=m\tanh{(nr)}$}
\begin{figure}
    \centering
    \includegraphics[width=8cm]{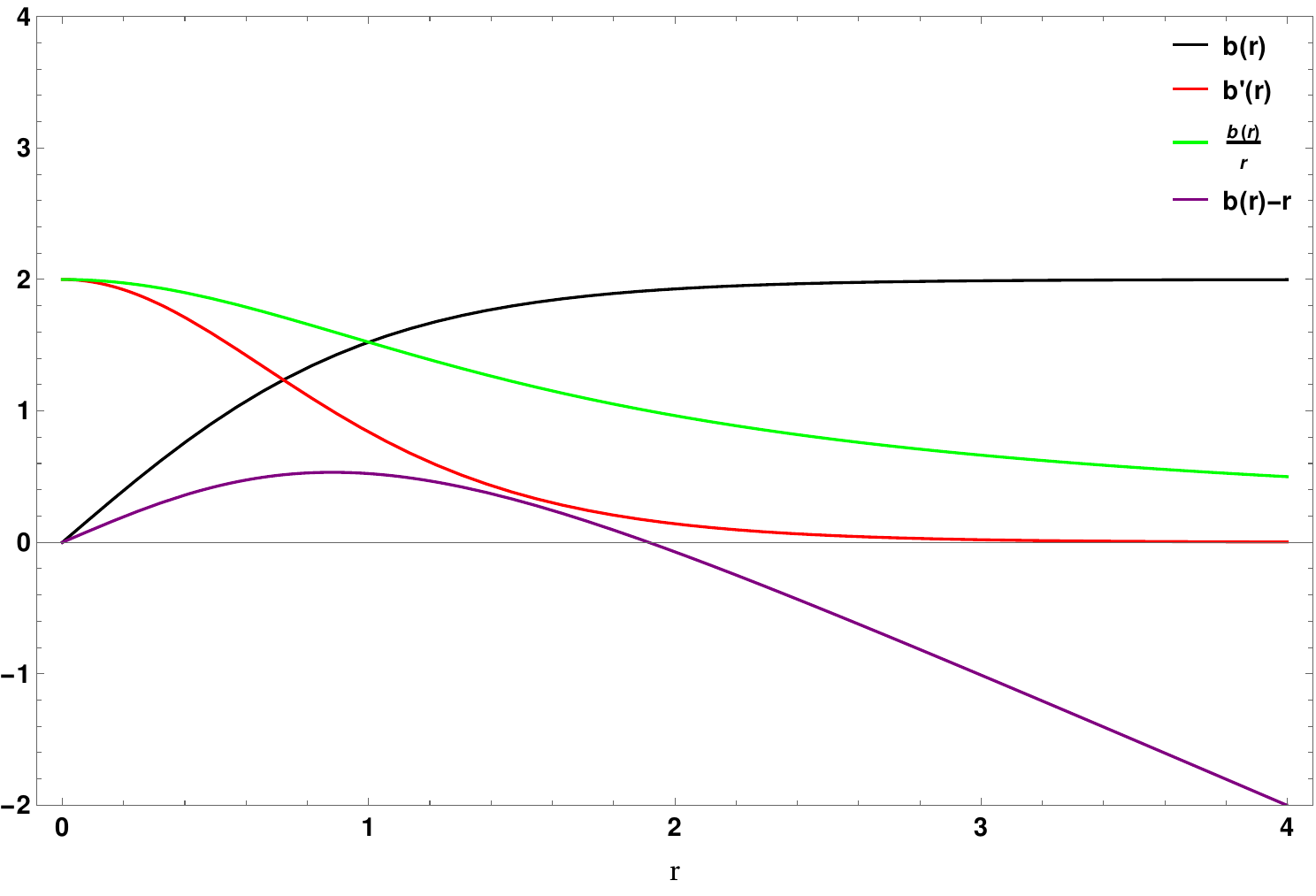}
    \caption{Plot showing the behaviour of shape function $b(r)=m \tanh{(nr)}$, throat condition, flaring out condition and asymptotically flatness condition for $m=2,n=1$}
    \label{b1}
\end{figure}

At first I have considered the redshift function as $\phi(r)=\frac{\phi_0}{r}$ and the shape function as, $b(r)=m \tanh{(nr)}$ where $\phi_0, m,n$ are arbitrary constants. This type of shape function for constant redshift has been analyzed in \cite{Rahaman3}. The shape function, flare out conditions and other conditions have been shown in Fig.~(\ref{b1}).
\\
From Eq.~(\ref{rho}), Eq.~(\ref{pr}) and Eq.~(\ref{pt}) I have derived the energy density, radial and lateral pressure for this particular choice of $\phi(r)$ and $b(r)$ as ,
\begin{equation}
    \rho=\frac{mnr^3(1+4\lambda)\sech^2{(nr)}+mr\lambda \sech{(nr)}(nr \sech{(nr)}-\sinh{(nr)})\phi_0+2\lambda\phi_0^2(r-m \tanh{(nr)})}{r^5(1+6\lambda+8\lambda^2)}
    \label{rho1}
\end{equation}
\begin{equation}
\begin{split}
   P_r &=-\frac{1}{r^5(1+6\lambda+8\lambda^2)}\bigl[mr^2(1+4\lambda)\tanh{(nr)}+2\lambda\phi_0^2(r-m \tanh{(nr)})+r\phi_0(2r(1+4\lambda)+\\
   &\quad mnr\lambda \sech^2{(nr)}-m(2+9\lambda)\tanh{(nr)})\bigr]
   \end{split}
   \label{pr1}
\end{equation}
\begin{equation}
\begin{split}
    P_t &=\frac{1}{2r^5(1+6\lambda+8\lambda^2)}\bigl[\frac{1}{2}mr^2(1+4\lambda)\sech^2{(nr)}(-2nr+\sinh{(2nr)})+2(1+2\lambda)\phi_0^2(r-m \tanh{(nr)})\\
    &\quad +r\phi_0(2r(1+4\lambda)+mnr(1+2\lambda)\sech^2{(nr)}-m \tanh{(nr)}(3+10\lambda))\bigr]
    \end{split}
    \label{pt1}
\end{equation}
The expression for energy conditions terms are calculated from Eq.~(\ref{rho1}-\ref{pt1}) as follows,
\begin{equation}
    \begin{split}
        \rho+P_r=\frac{mr(nr \sech^2{(nr)}-\tanh{(nr)})-2\phi_0(r-m\tanh{(nr)})}{r^4(1+2\lambda)}
    \end{split}
\end{equation}
\begin{equation}
    \begin{split}
        \rho+P_t  &=\frac{1}{2r^5(1+2\lambda)}\bigl[mr^2(nr \sech^2{(nr)}+\tanh{(nr)})+r\phi_0(2r+mnr \sech^2{(nr)}-3m\tanh{(nr)})\\
        &\quad +2\phi_0^2(r-m\tanh{(nr)})\bigr]
    \end{split}
\end{equation}
\begin{equation}
    \begin{split}
        \rho+P_r+2P_t &=\frac{\sech^2{(nr)}\phi_0(mr(2nr-\sinh{(2nr)})+2(r+r \cosh{(2nr)}-m \sinh{(2nr)})\phi_0)}{2r^5(1+4\lambda)}
    \end{split}
\end{equation}
\begin{equation}
    \begin{split}
        \rho-P_r &=\frac{1}{r^5(1+6\lambda+8\lambda^2)}\bigl[mr^2(1+4\lambda)\sech{(nr)}(nr \sech{(nr)}+\sinh{(nr)})+4\lambda\phi_0^2(r-m\tanh{(nr)})\\
        &\quad +2r\phi_0(r+4r\lambda+mnr\lambda \sech^2{(nr)}-m(1+15\lambda)\tanh{(nr)}\bigr]
    \end{split}
\end{equation}
\begin{equation}
    \begin{split}
        \rho-P_t &=\frac{1}{2r^5(1+6\lambda+8\lambda^2)}\bigl[mr^2(1+4\lambda)\sech{(nr)}(3nr\sech{(nr)}-\sinh{(nr)})-2\phi_0^2(r-m \tanh{(nr)})\\
        & \quad -r\phi_0(2r(1+4\lambda)+mnr \sech^2{(nr)}-m(3+8\lambda)\tanh{(nr)})\bigr]
    \end{split}
\end{equation}

\begin{figure}
    \centering
    \includegraphics[width=5cm]{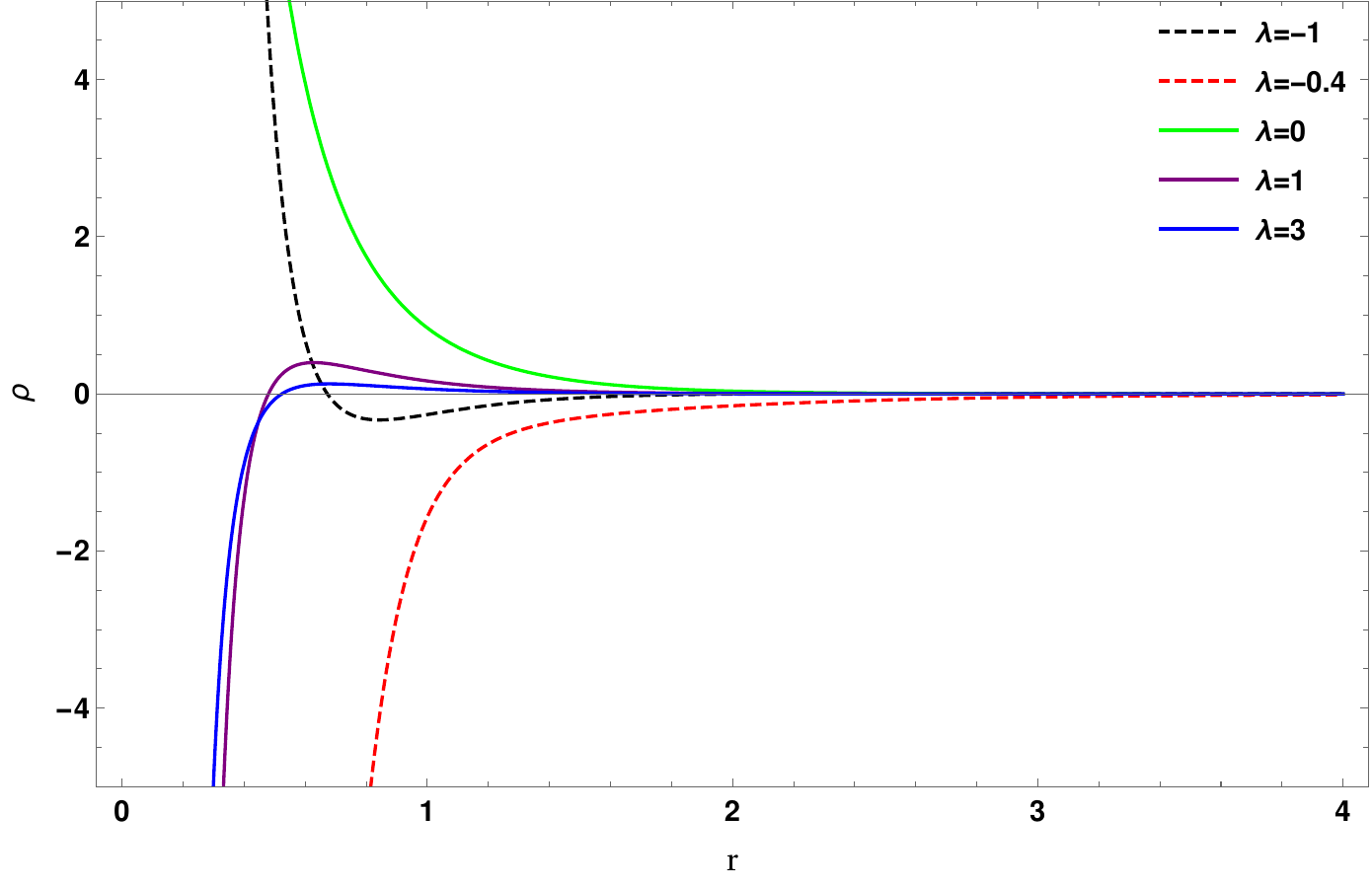}
    \vspace*{0.1in}
    \includegraphics[width=5cm]{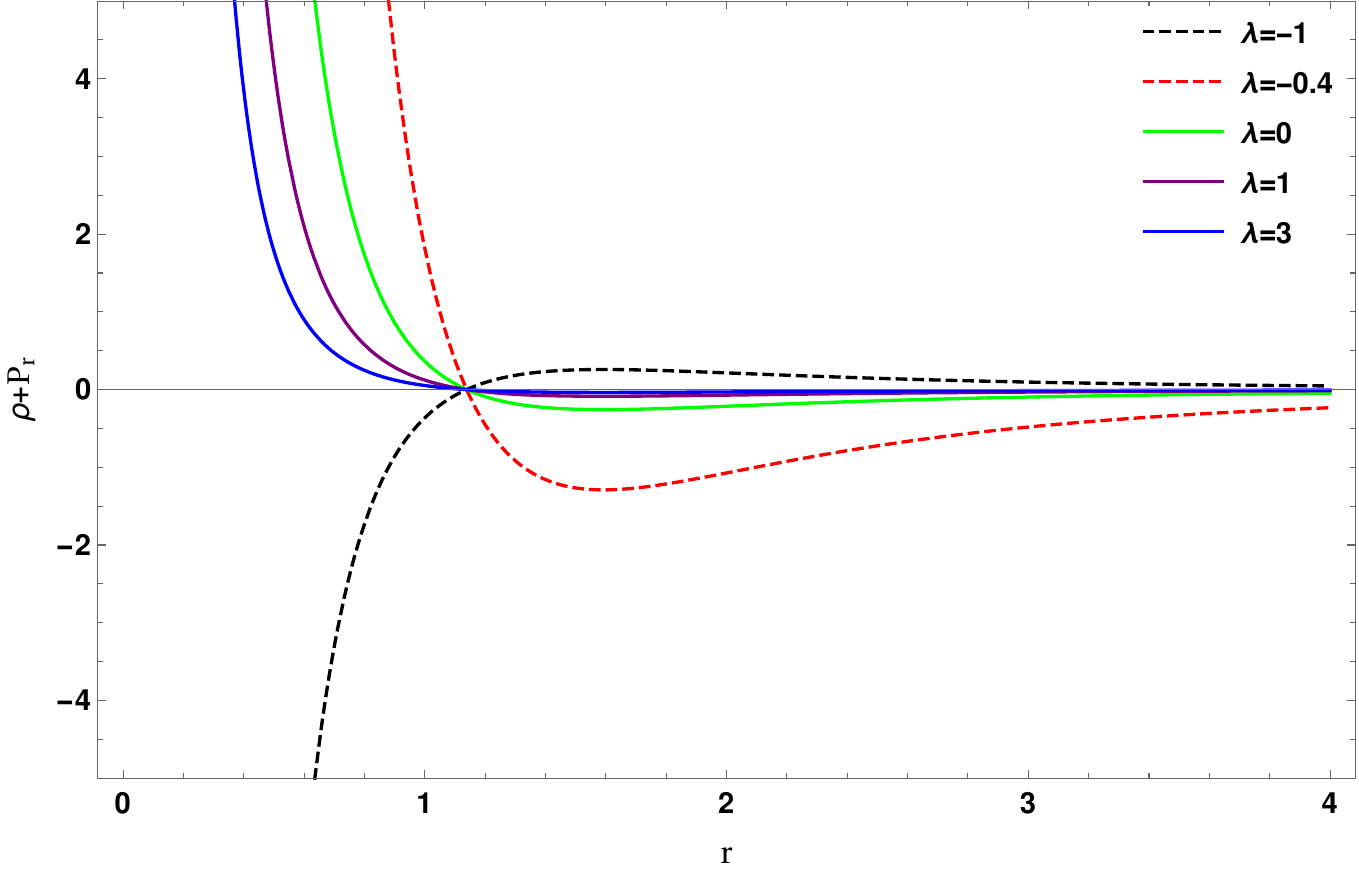}
    \vspace*{0.1in}
    \includegraphics[width=5cm]{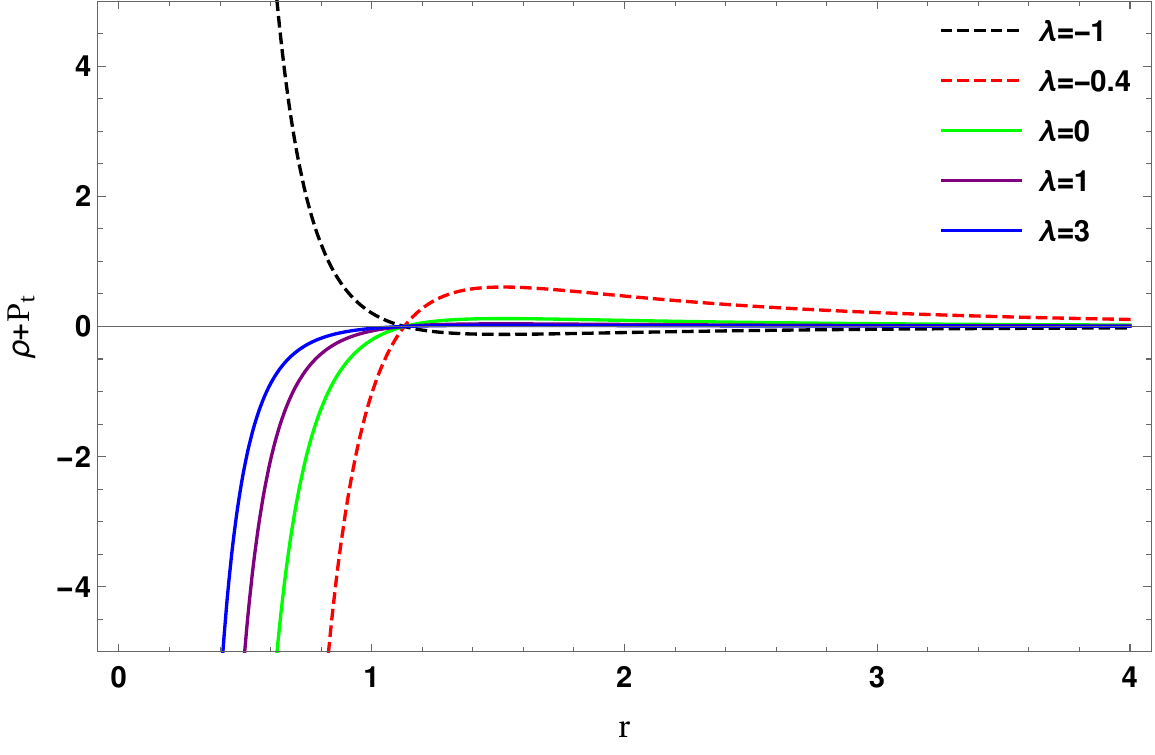}
    \vspace*{0.1in}
    \includegraphics[width=5cm]{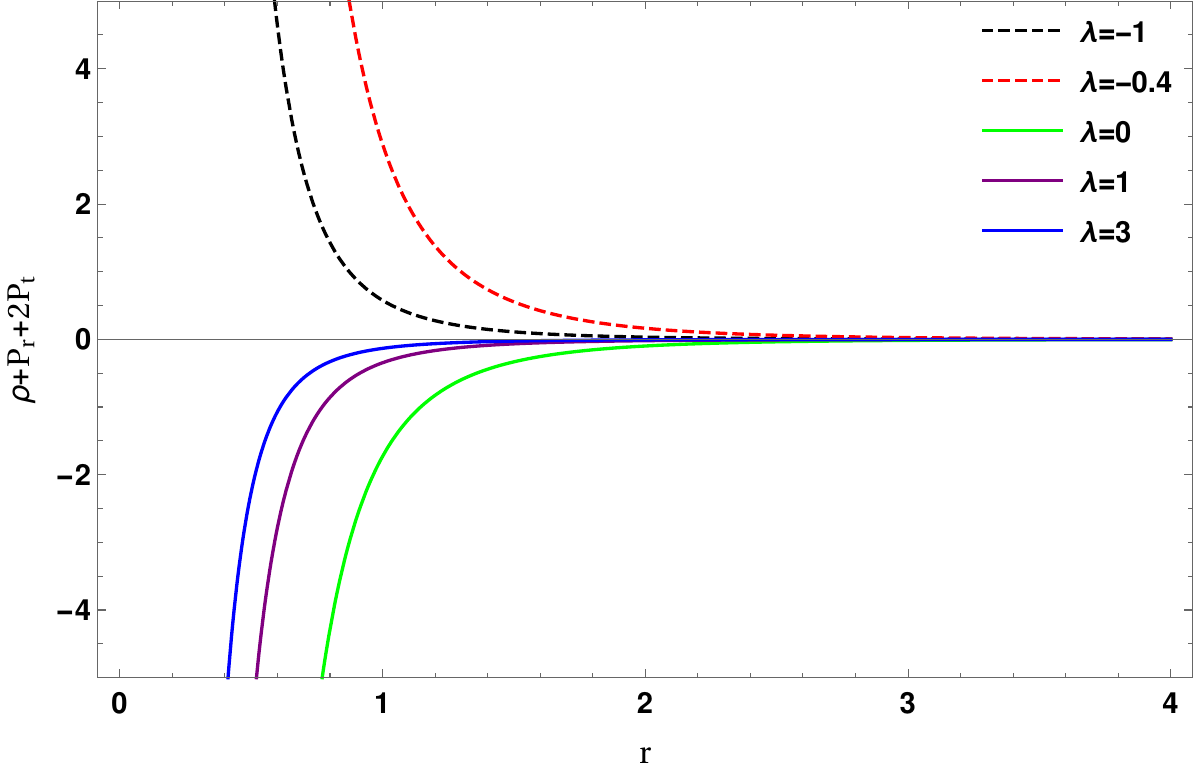}
    \vspace*{0.1in}
    \includegraphics[width=5cm]{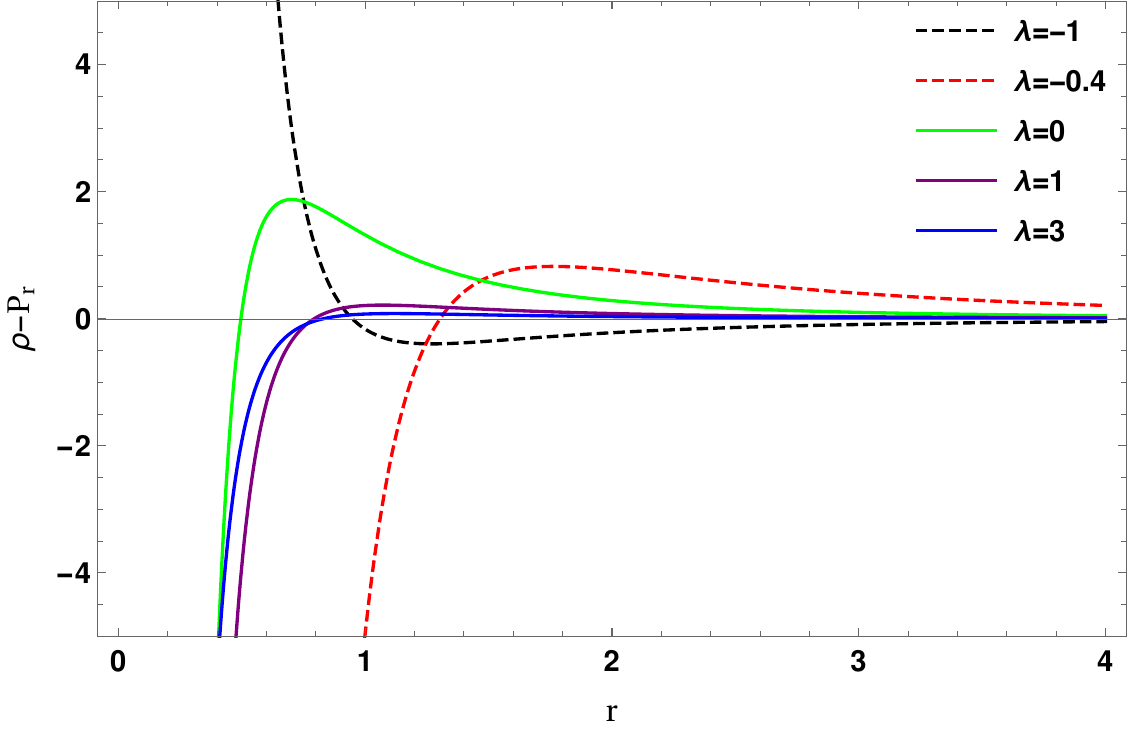}
    \vspace*{0.1in}
    \includegraphics[width=5cm]{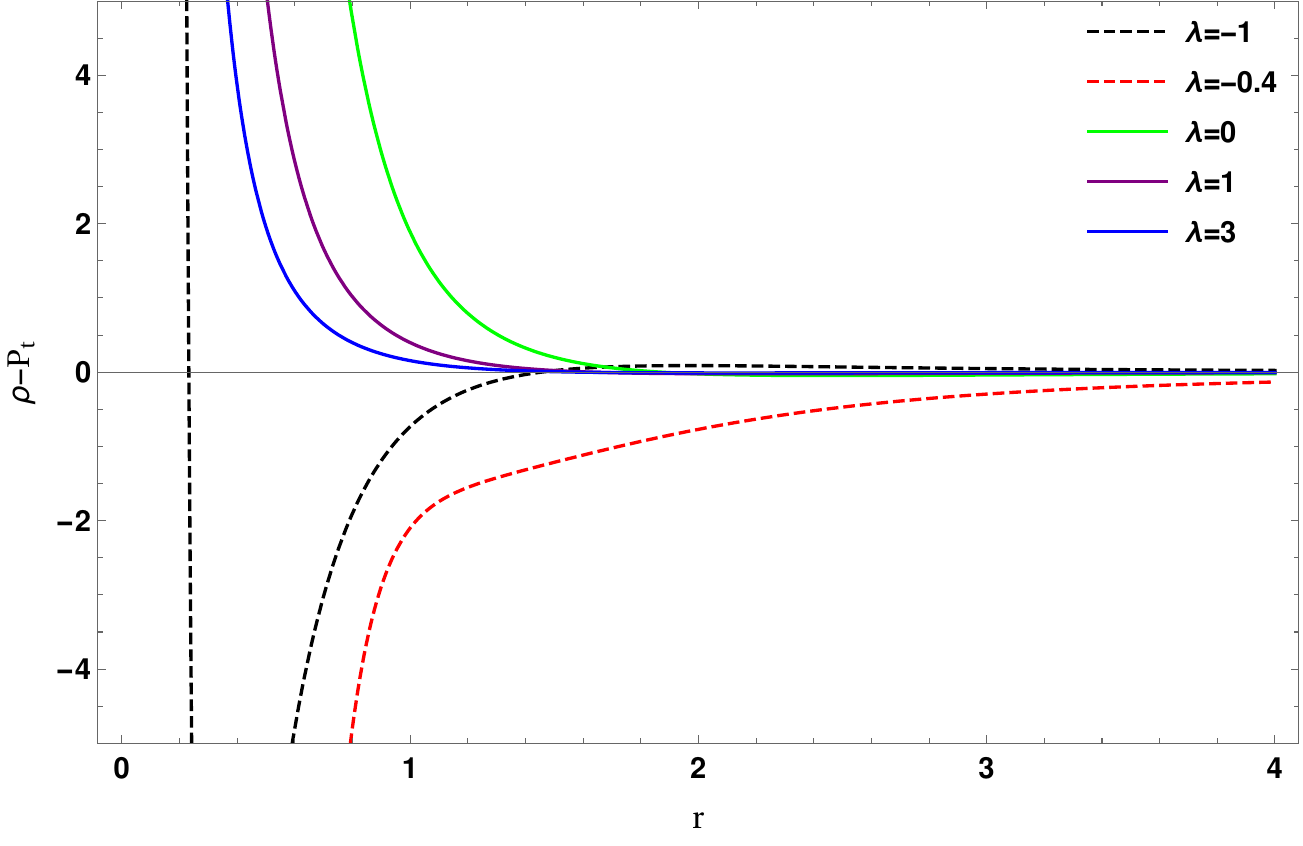}
    \caption{Plots for energy conditions terms with respect to $r$ for different $\lambda$ values}
    \label{energyconditions1}
\end{figure}
\subsection{Case 2: $\phi(r)=\frac{\phi_0}{r}$, $b(r)=r_0+ar_0\left(\frac{1}{r}-\frac{1}{r_0}\right)$ }
\begin{figure}
    \centering
    \includegraphics[width=8cm]{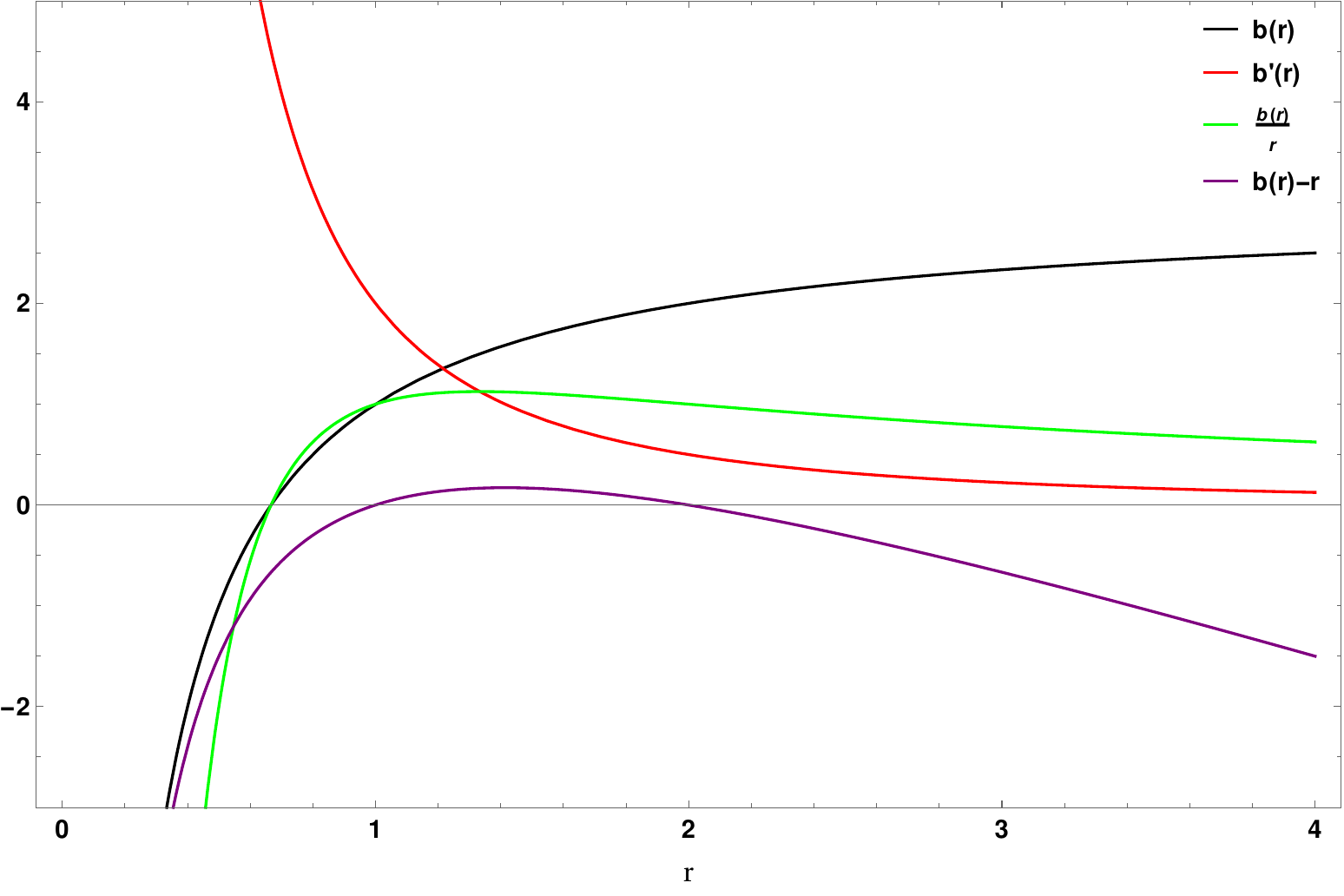}
    \caption{Plot showing the behaviour of shape function $b(r)=r_0+ar_0\left(\frac{1}{r}-\frac{1}{r_0}\right)$, throat condition, flaring out condition and asymptotically flatness condition for $a=-1,r_0=2$}
    \label{b2}
\end{figure}

In this section I have taken $b(r)=r_0+ar_0\left(\frac{1}{r}-\frac{1}{r_0}\right)$ for same redshift function. In \cite{Mishra4} has discussed this particular shape function in $f(R)$ gravity with constant redshift function. The behaviour of the shape function along with throat conditions have shown in Fig.~(\ref{b2}).
\\
The energy density and pressures can be calculated from Eq.~(\ref{rho}-\ref{pt}) as before,
\begin{equation}
    \rho=\frac{-ar_0 r^2(1+4\lambda)+r(-rr_0+a(r-2r_0))\lambda\phi_0-2\lambda\phi_0^2(a+r)(r_0-r)}{r^6(1+6\lambda+8\lambda^2)}
\end{equation}
\begin{equation}
    \begin{split}
        P_r &=\frac{1}{r^6(1+6\lambda+8\lambda^2)}\bigl(r^2(-rr_0+a(-r_0+r))(1+4\lambda)+r(a(2r_0-2r+10\lambda r_0-9r\lambda)\\
        &\quad +r(2r_0-2r+9r_0\lambda-8r\lambda))\phi_0+2\lambda\phi_0^2(r_0-r)(a+r)\bigr)
    \end{split}
\end{equation}

\begin{equation}
\begin{split}
P_t &=\frac{1}{2r^6(1+6\lambda+8\lambda^2)}\bigl(r^2(2ar_0-ar+rr_0)(1+4\lambda)+r(r(-3r_0+2r-10\lambda r_0+8r\lambda)\\
&\quad +a(-4r_0(1+3\lambda)+r(3+10\lambda)))\phi_0-2(r_0-r)(a+r)\phi_0^2(1+2\lambda)\bigr)
\end{split}
\end{equation}
\begin{figure}
    \centering
    \includegraphics[width=5cm]{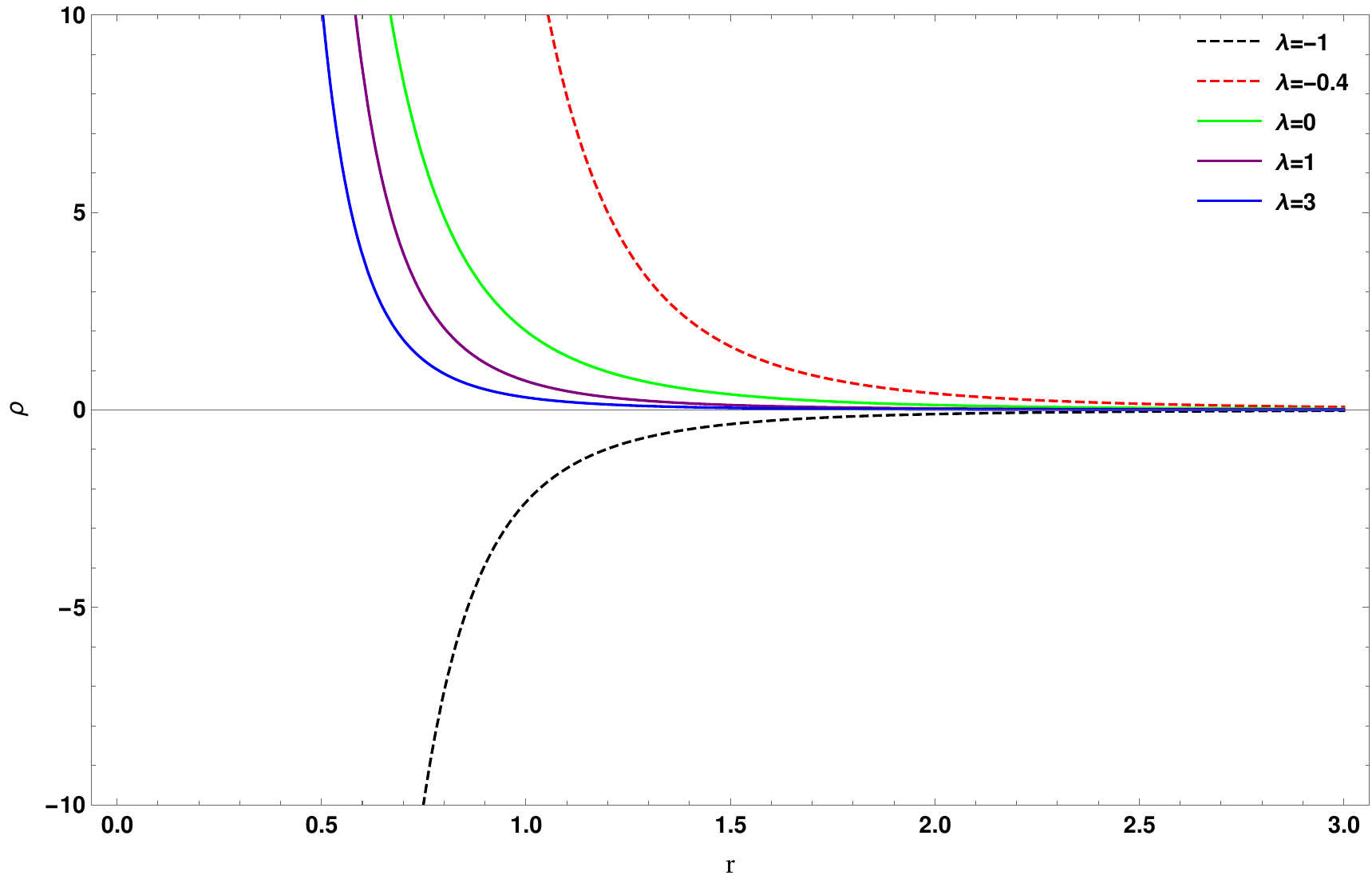}
    \vspace*{0.1in}
    \includegraphics[width=5cm]{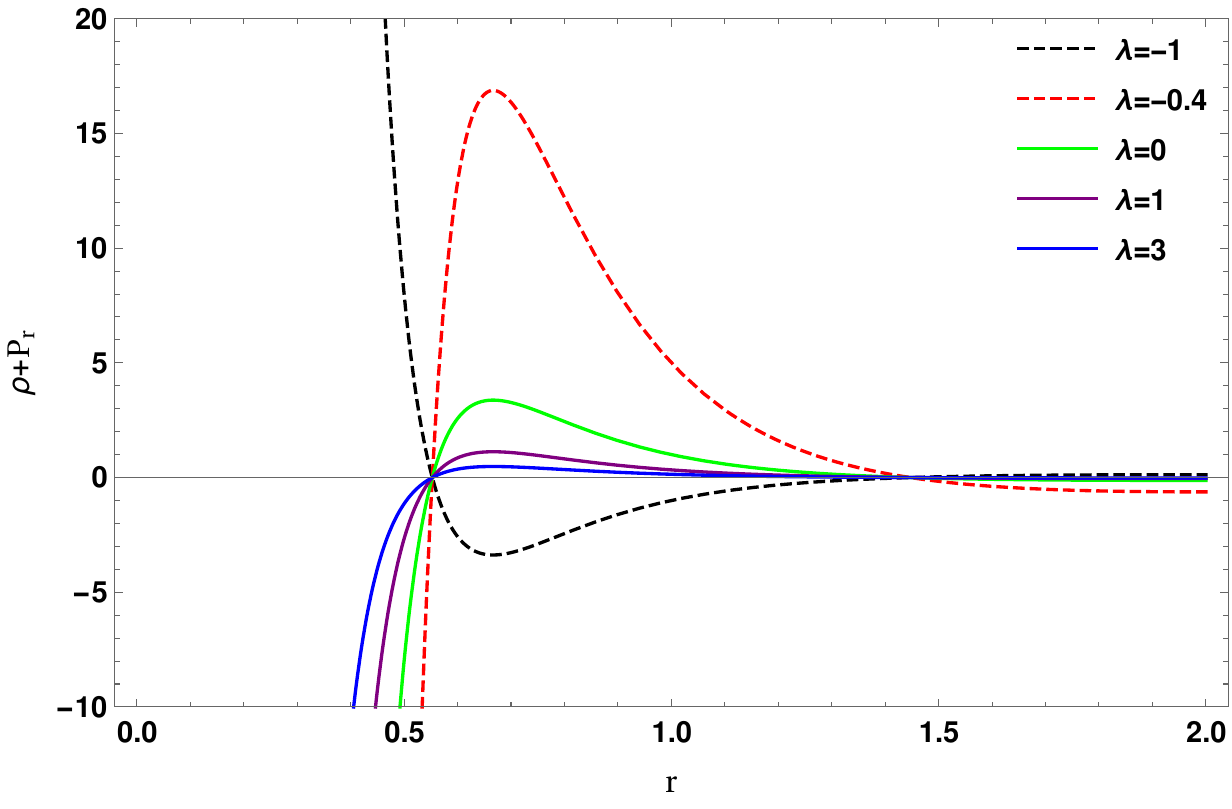}
    \vspace*{0.1in}
    \includegraphics[width=5cm]{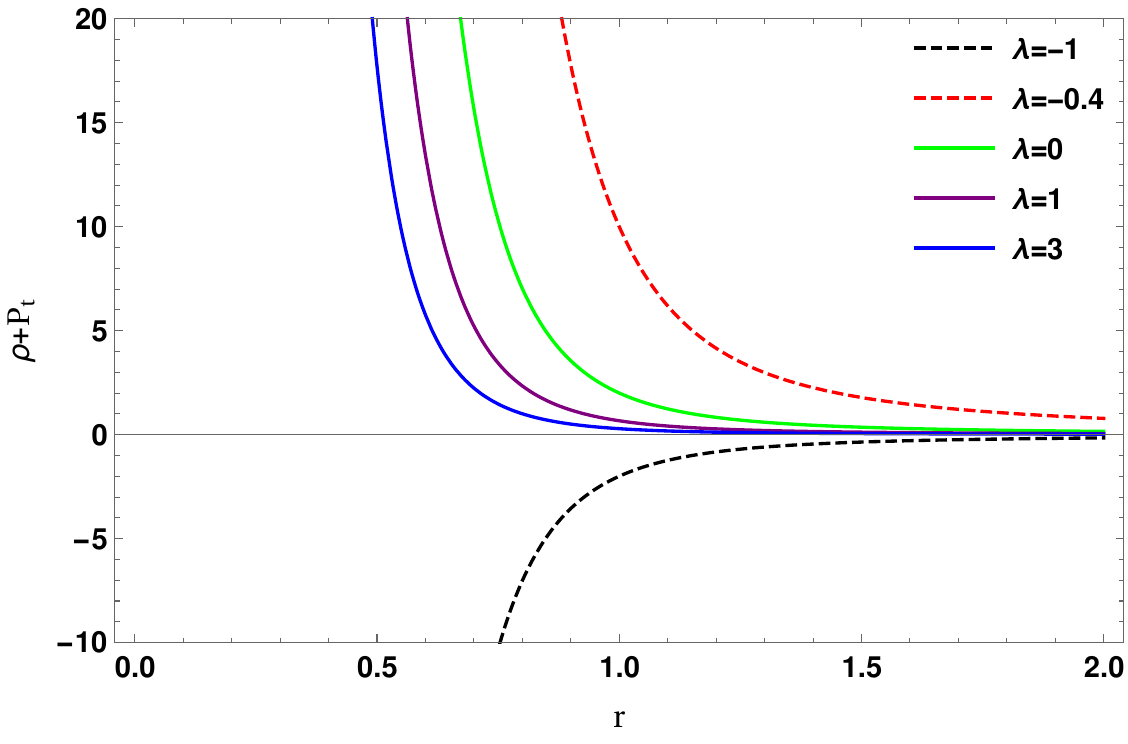}
    \vspace*{0.1in}
    \includegraphics[width=5cm]{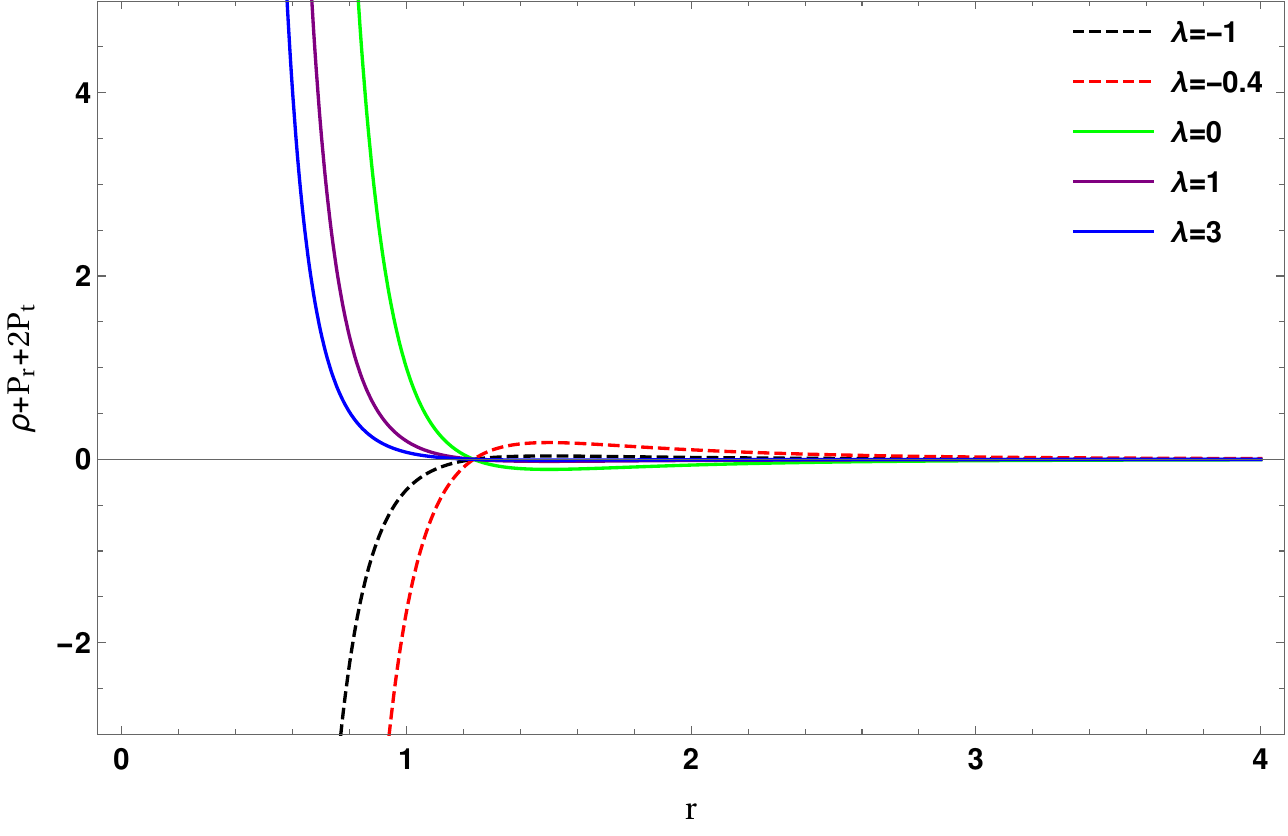}
    \vspace*{0.1in}
    \includegraphics[width=5cm]{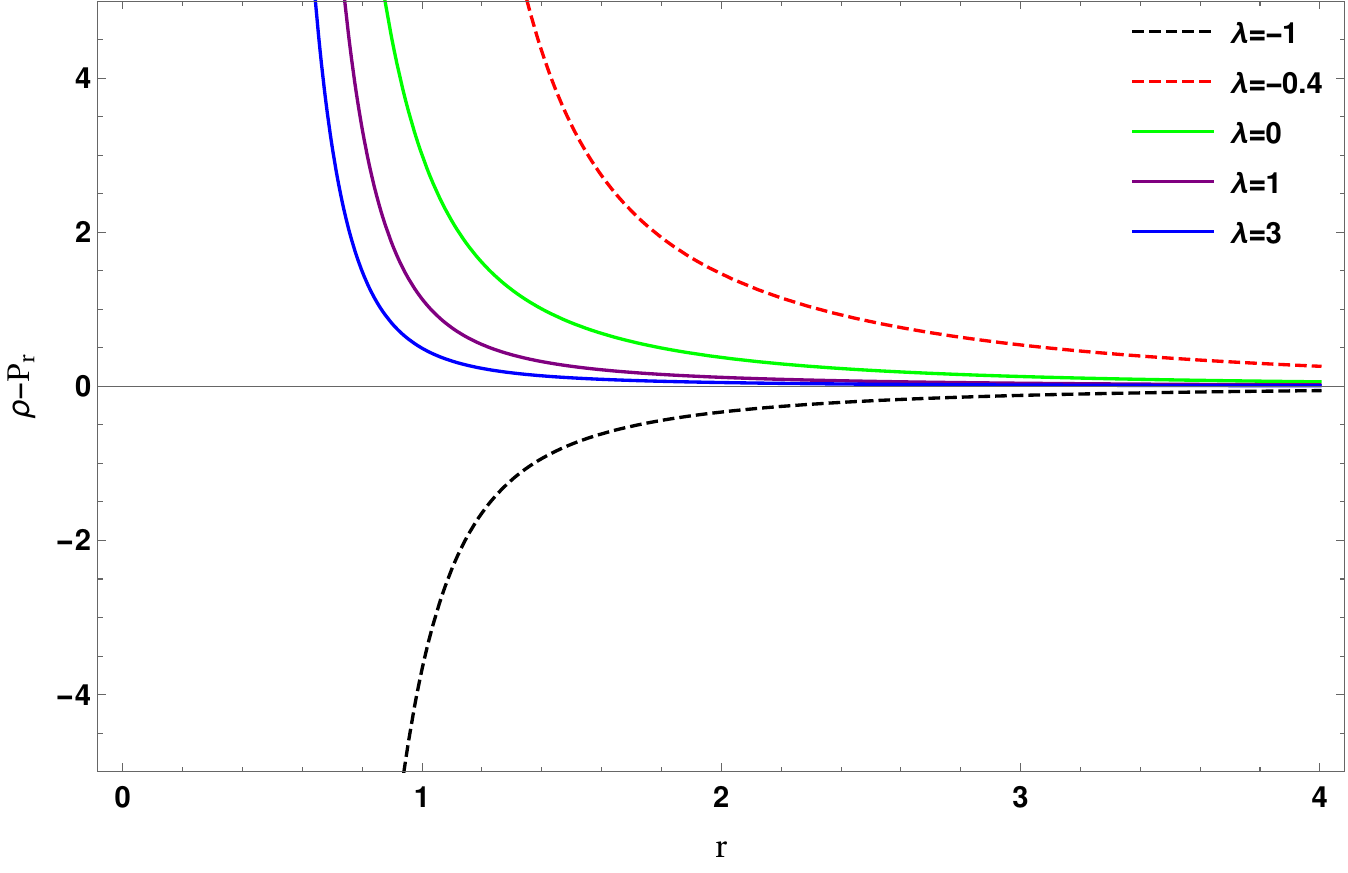}
    \vspace*{0.1in}
    \includegraphics[width=5cm]{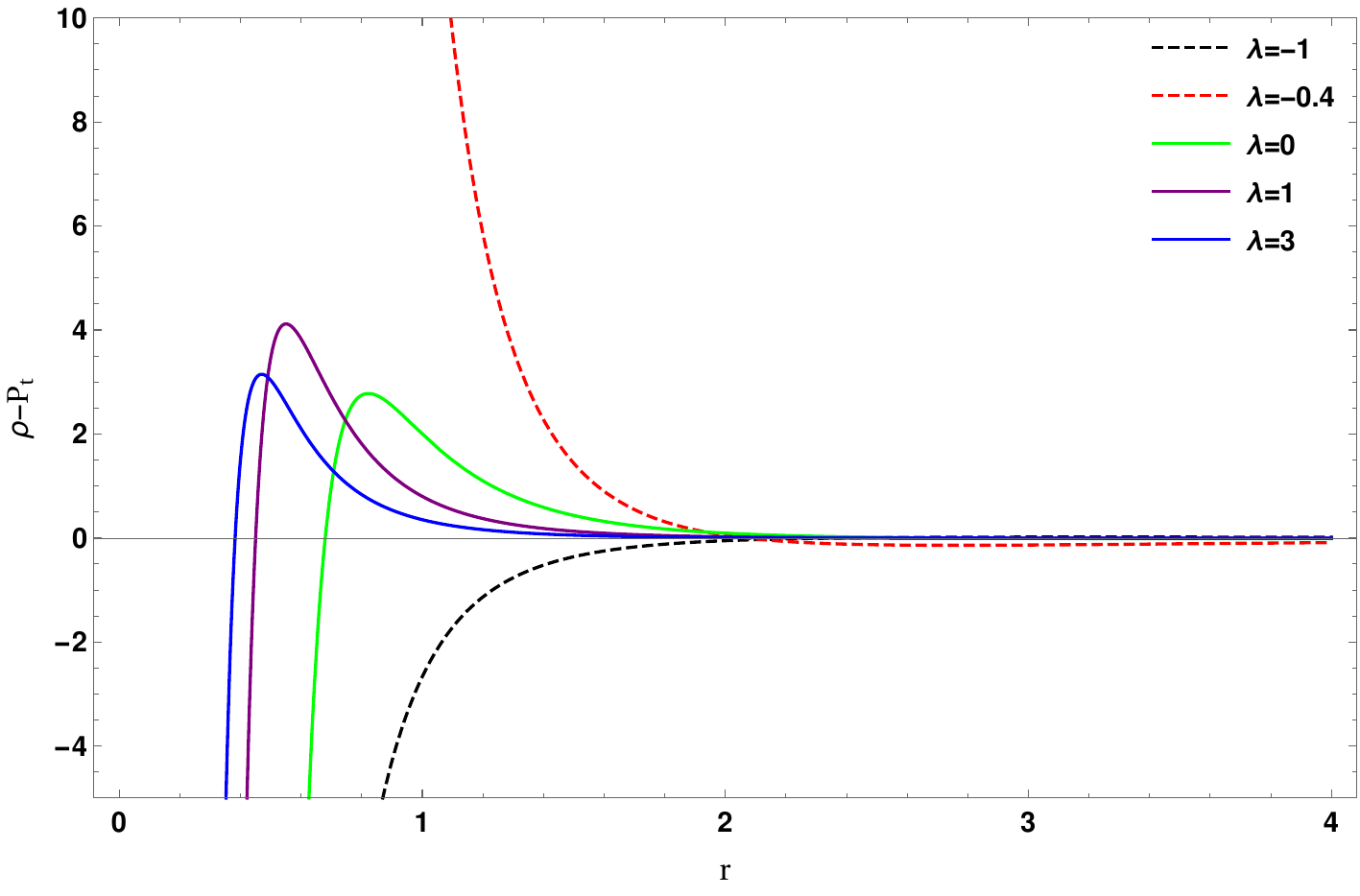}
    \caption{Plots for energy conditions terms with respect to $r$ for different $\lambda$ values}
    \label{energyconditions2}
\end{figure}
Now I will follow the same path as before. I will study the energy conditions after calculating $\rho+|P_r|, \rho+|P_t|, \rho+P_r+2P_t$ etc.
The expression for $\rho+P_r$ will be as,
\begin{equation}
    \rho+P_r=\frac{r(-rr_0+a(r-2r_0))+2\phi_0(a+r)(r_0-r)}{r^5(1+2\lambda)}
\end{equation}
Similarly,
\begin{equation}
    \rho+P_t=\frac{r^3(-a+r_0)+r\phi_0(-4ar_0+3ar-3rr_0+2r^2)-2\phi_0^2(r_0-r)(a+r)}{2r^6(1+2\lambda)}
\end{equation}

\begin{equation}
    \rho+P_r+2P_t=\frac{\phi_0(r(-rr_0+a(-2r_0+r)-2(r_0-r)(a+r)\phi_0)}{r^6(1+4\lambda)}
\end{equation}
\begin{equation}
    \begin{split}
        \rho-P_r &=-\frac{1}{r^6(1+6\lambda+8\lambda^2)}\bigl((a-r_0)r^3(1+4\lambda)+2r(r(r_0+5r_0\lambda-r(1+4\lambda))+a(r_0+6r_0\lambda\\
        &\quad -r(1+5\lambda))\phi_0+4\lambda\phi_0^2(a+r)(r_0-r)\bigr)
    \end{split}
\end{equation}
\begin{equation}
    \begin{split}
        \rho-P_t &=\frac{1}{2r^6(1+6\lambda+8\lambda^2)}\bigl(r^2(-rr_0+a(r-4r_0))(1+4\lambda)+r(a(4r_0-3r+8r_0\lambda-8r\lambda)\\
        &\quad +r(3r_0-2r+8r_0\lambda-8r\lambda))\phi_0+2\phi_0^2(a+r)(r_0-r)\bigr)
    \end{split}
\end{equation}

\subsection{Case 3: $\phi(r)=\frac{\phi_0}{r}$, $b(r)=r_0\left(\frac{a^r}{a^{r_0}}\right)$}
Next, I have considered here the shape function as $b(r)=r_0\left(\frac{a^r}{a^{r_0}}\right)$ with the same redshift function as mentioned before.
\begin{figure}
    \centering
    \includegraphics[width=8cm]{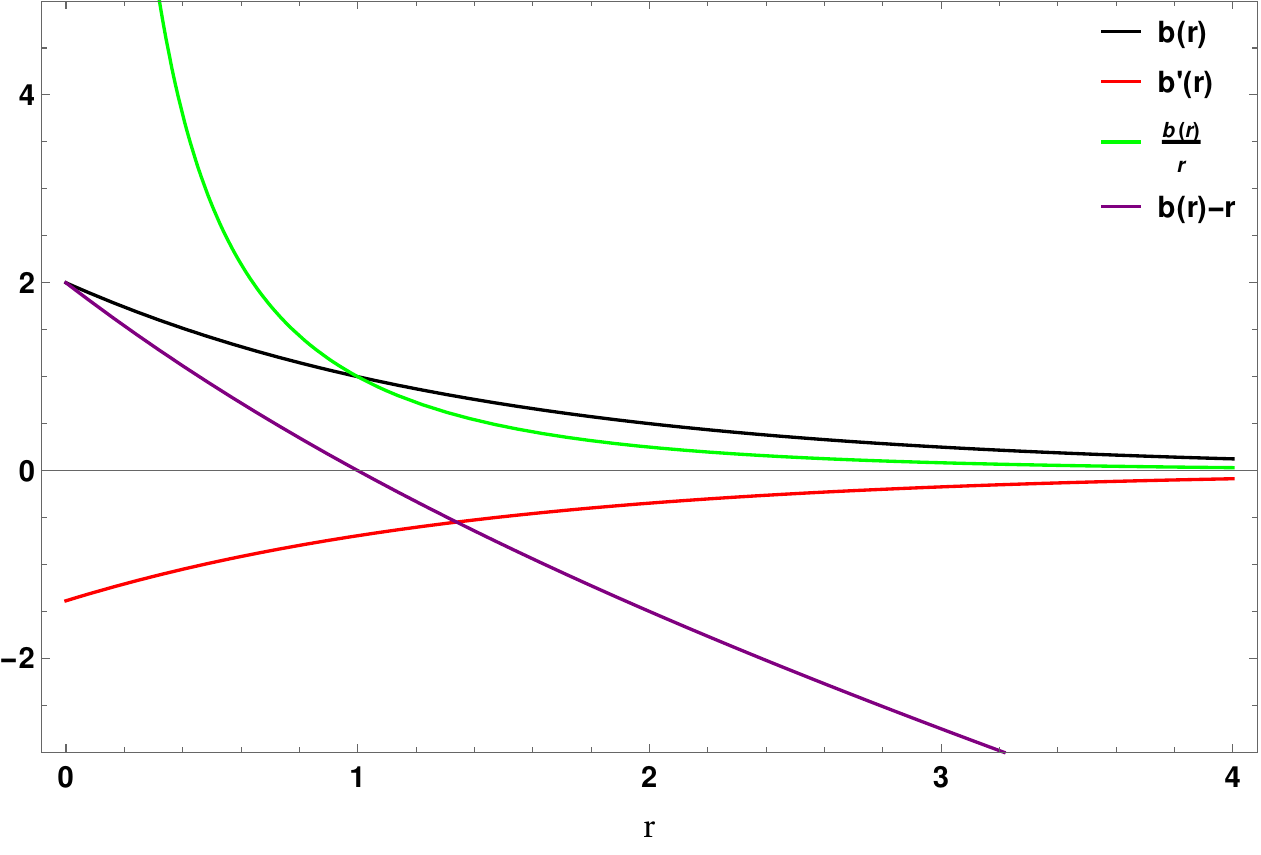}
    \caption{Plot showing the behaviour of shape function $b(r)=r_0\left(\frac{a^r}{a^r_0}\right)$, throat condition, flaring out condition and asymptotically flatness condition for $a=1/2,r_0=1$}
    \label{b3}
\end{figure}
The energy density and pressures will be,
\begin{equation}
    \rho=\frac{a^{-r_0}(a^rr_0r^3\ln{(a)}(1+4\lambda)+a^rr_0r\lambda(-1+r\ln{(a)})\phi_0+2\lambda\phi_0^2(-a^rr_0+ra^{r_0}))}{r^5(1+6\lambda+8\lambda^2)}
    \label{rho3}
\end{equation}
\begin{equation}
\begin{split}
    P_r &=\frac{1}{r^5(1+6\lambda+8\lambda^2)}\bigl(a^{-r_0}(-a^rr_0r^2(1+4\lambda)-r(2ra^{r_0}(1+4\lambda)-a^rr_0(2+9\lambda)\\
    &\quad +a^rrr_0\lambda \ln{(a)})\phi_0-2\phi_0^2\lambda(ra^{r_0}-a^rr_0))\bigr)
\end{split}
\label{pr3}
\end{equation}
\begin{equation}
    \begin{split}
      P_t &=\frac{1}{2r^5(1+6\lambda+8\lambda^2)}\bigl(a^{-r_0}(-a^rr_0r^2(1+4\lambda)(-1+r\ln{(a)})+r(2ra^{r_0}(1+4\lambda)\\
      &\quad-r_0a^r(3+10\lambda)+r_0ra^r\ln{(a)}(1+2\lambda))\phi_0+2\phi_0^2(1+2\lambda)(ra^{r_0}-r_0a^r))\bigr)  
    \end{split}
    \label{pt3}
\end{equation}
\begin{figure}
    \centering
    \includegraphics[width=5cm]{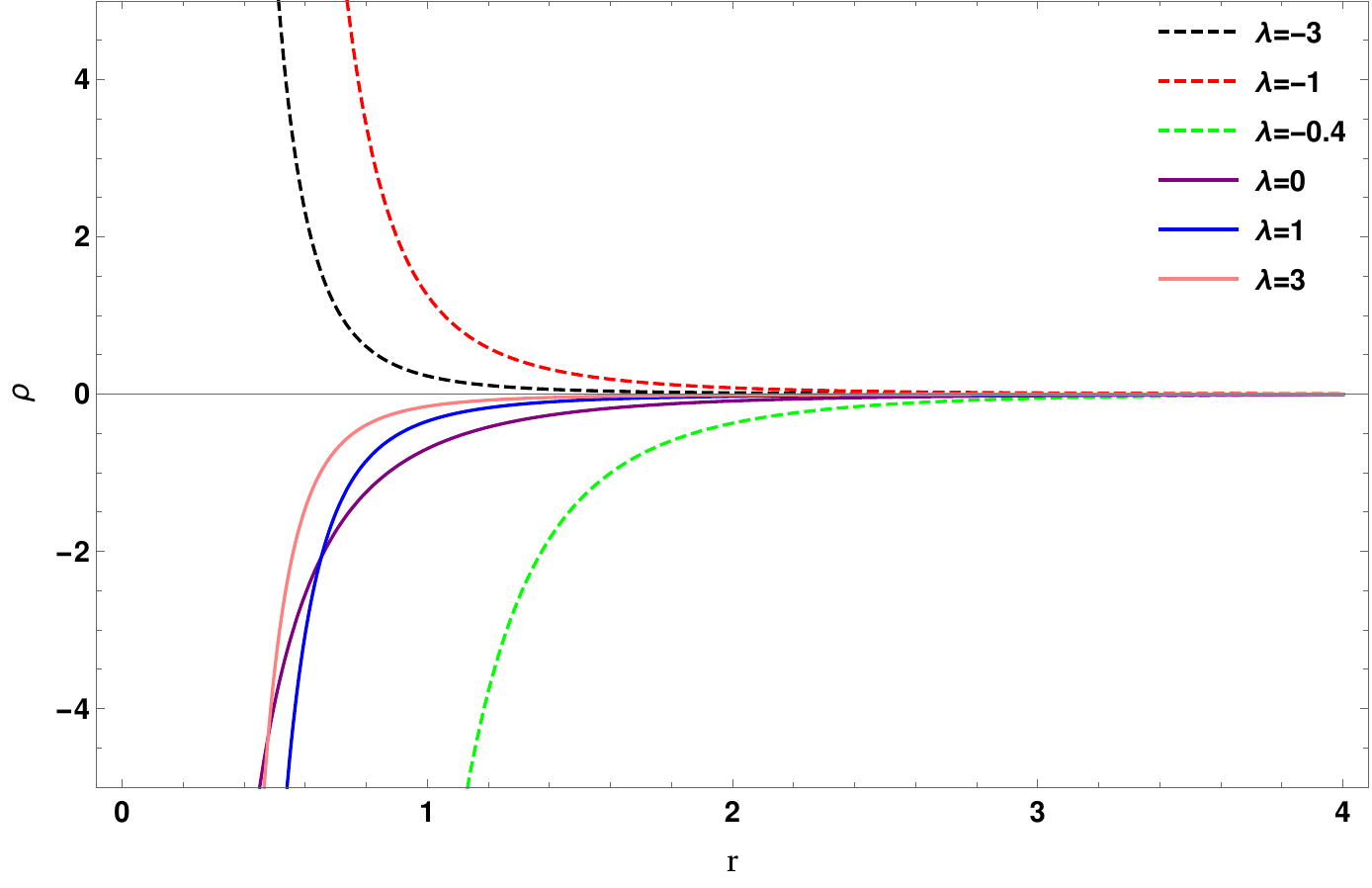}
    \vspace*{0.1in}
    \includegraphics[width=5cm]{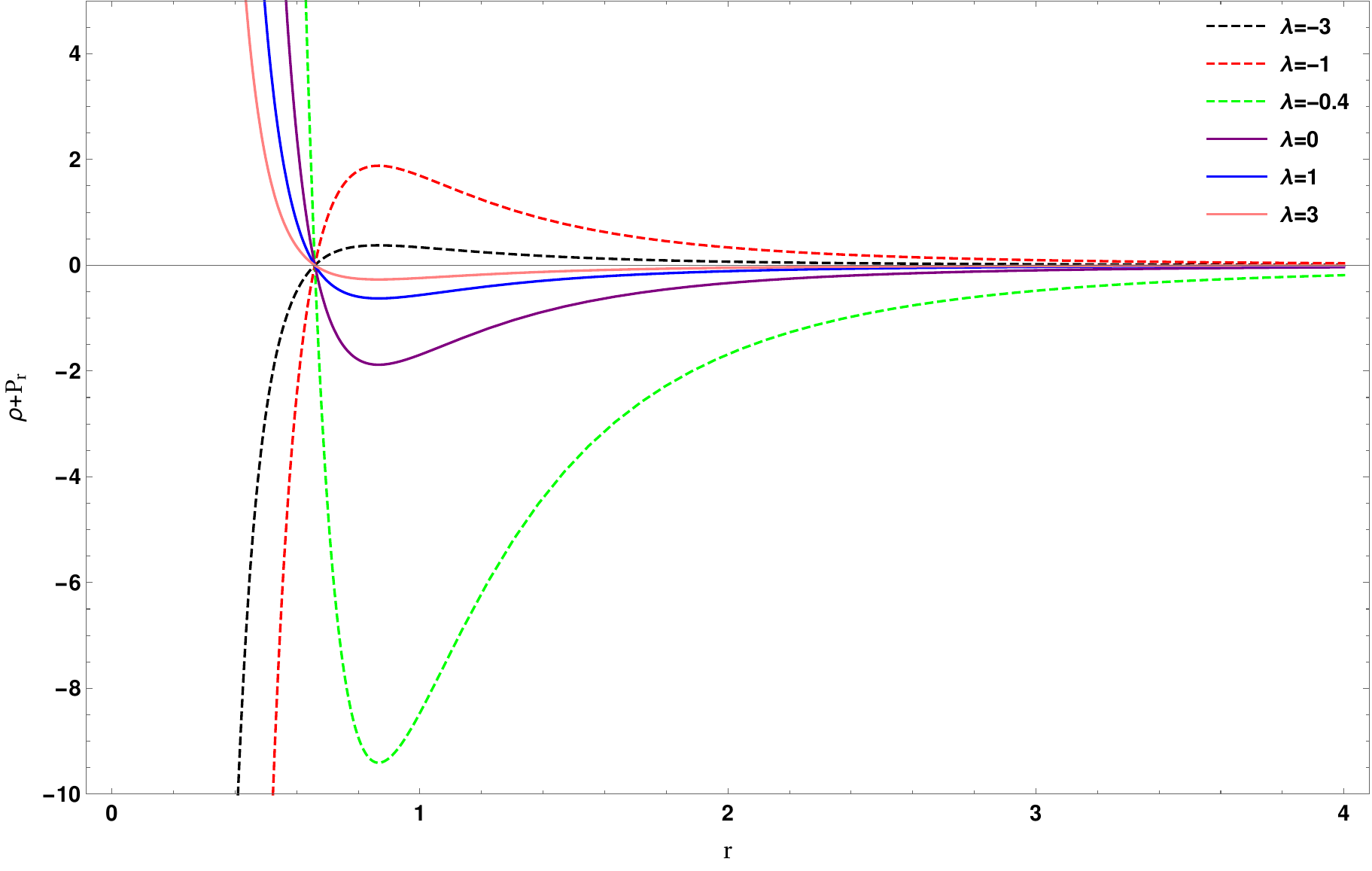}
    \vspace*{0.1in}
    \includegraphics[width=5cm]{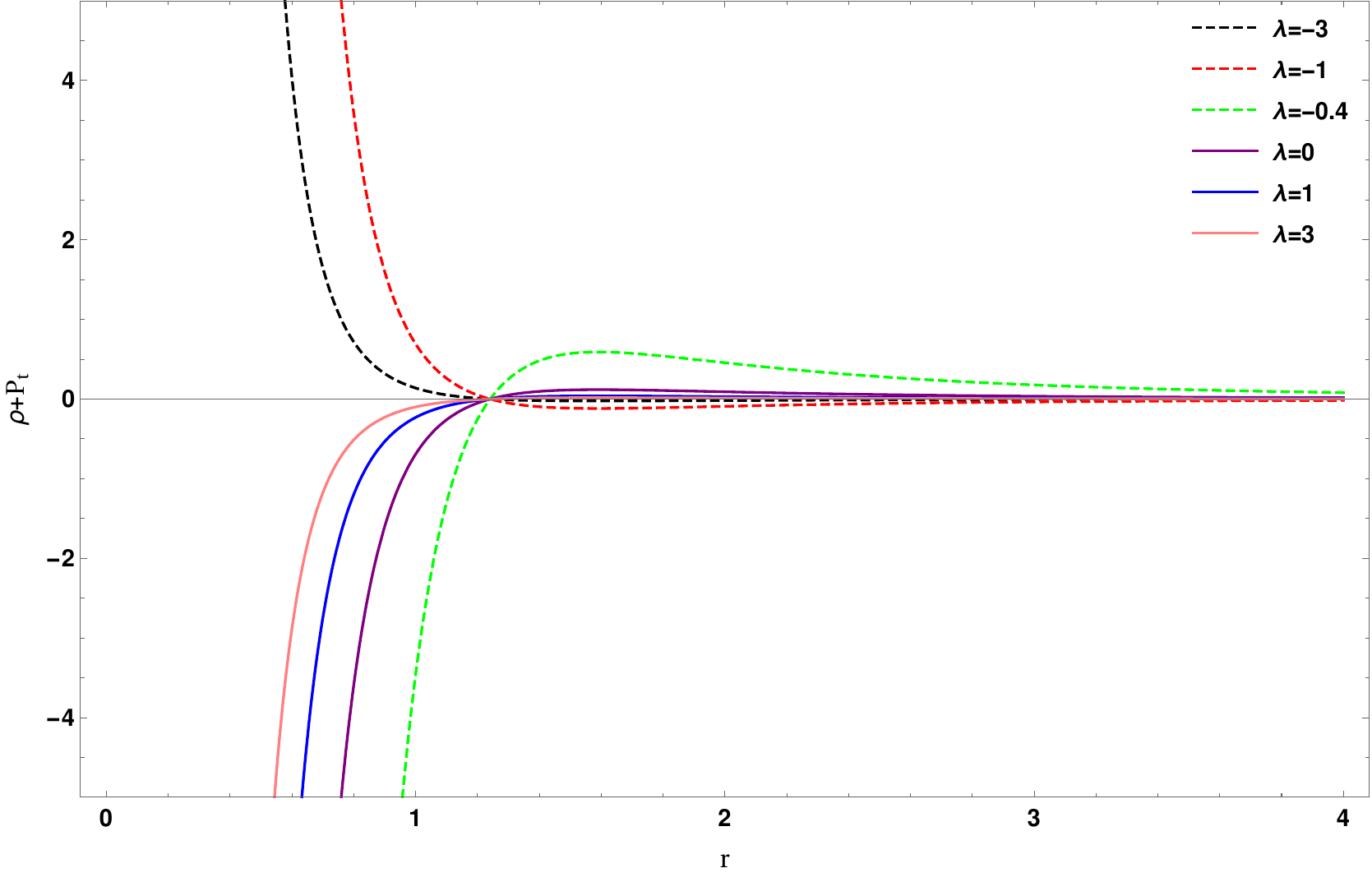}
    \vspace*{0.1in}
    \includegraphics[width=5cm]{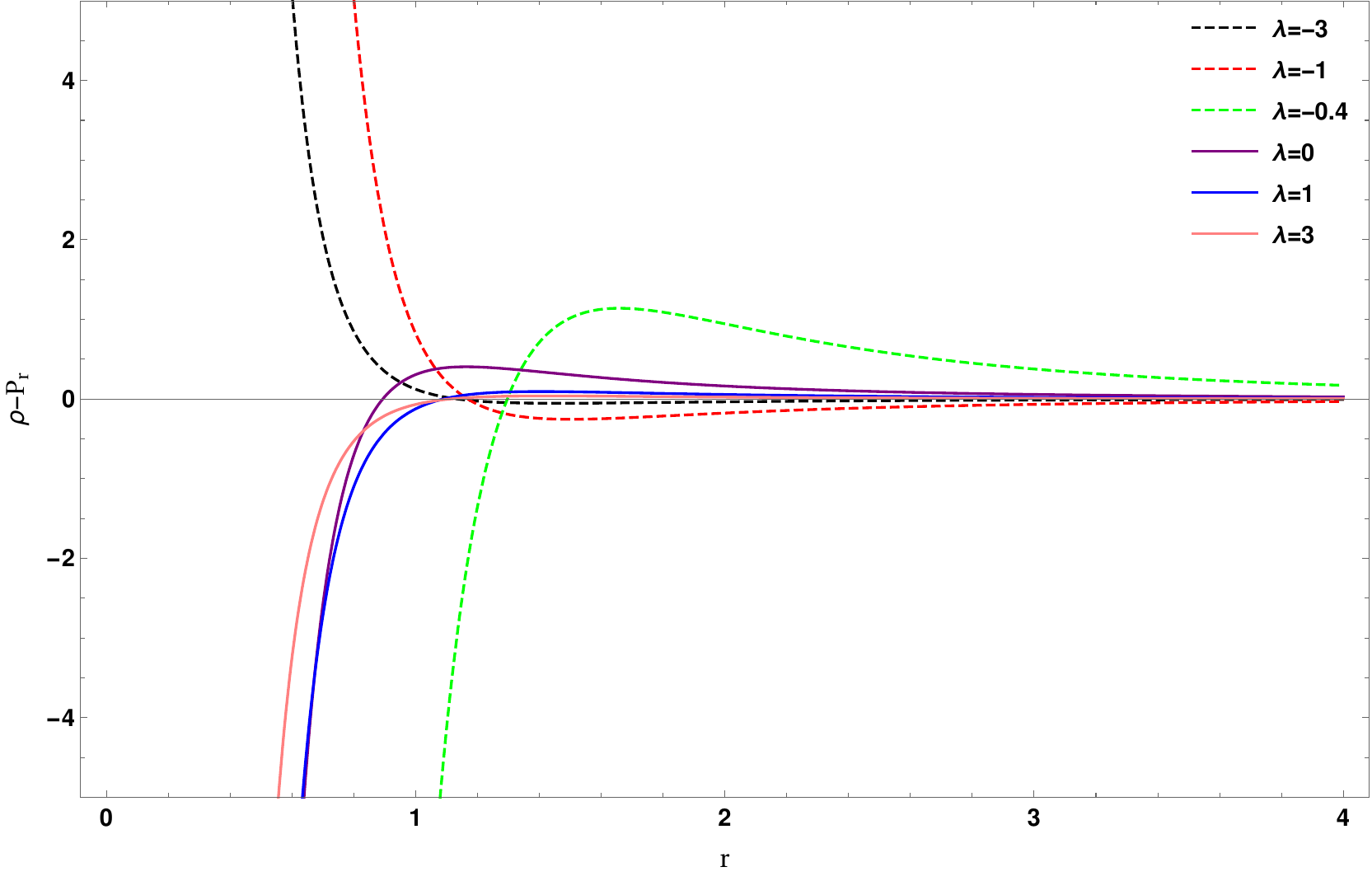}
    \vspace*{0.1in}
    \includegraphics[width=5cm]{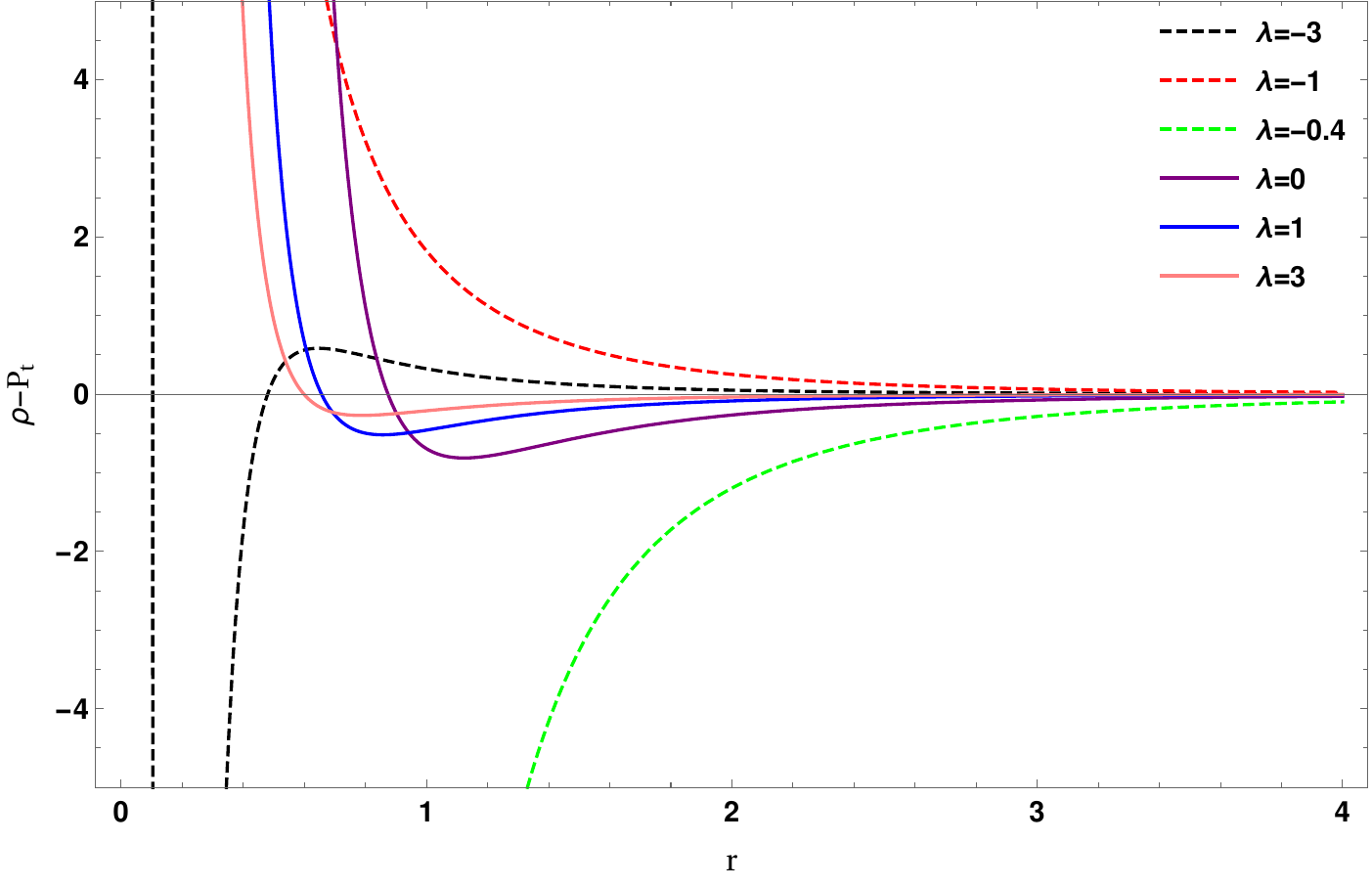}
    \vspace*{0.1in}
    \includegraphics[width=5cm]{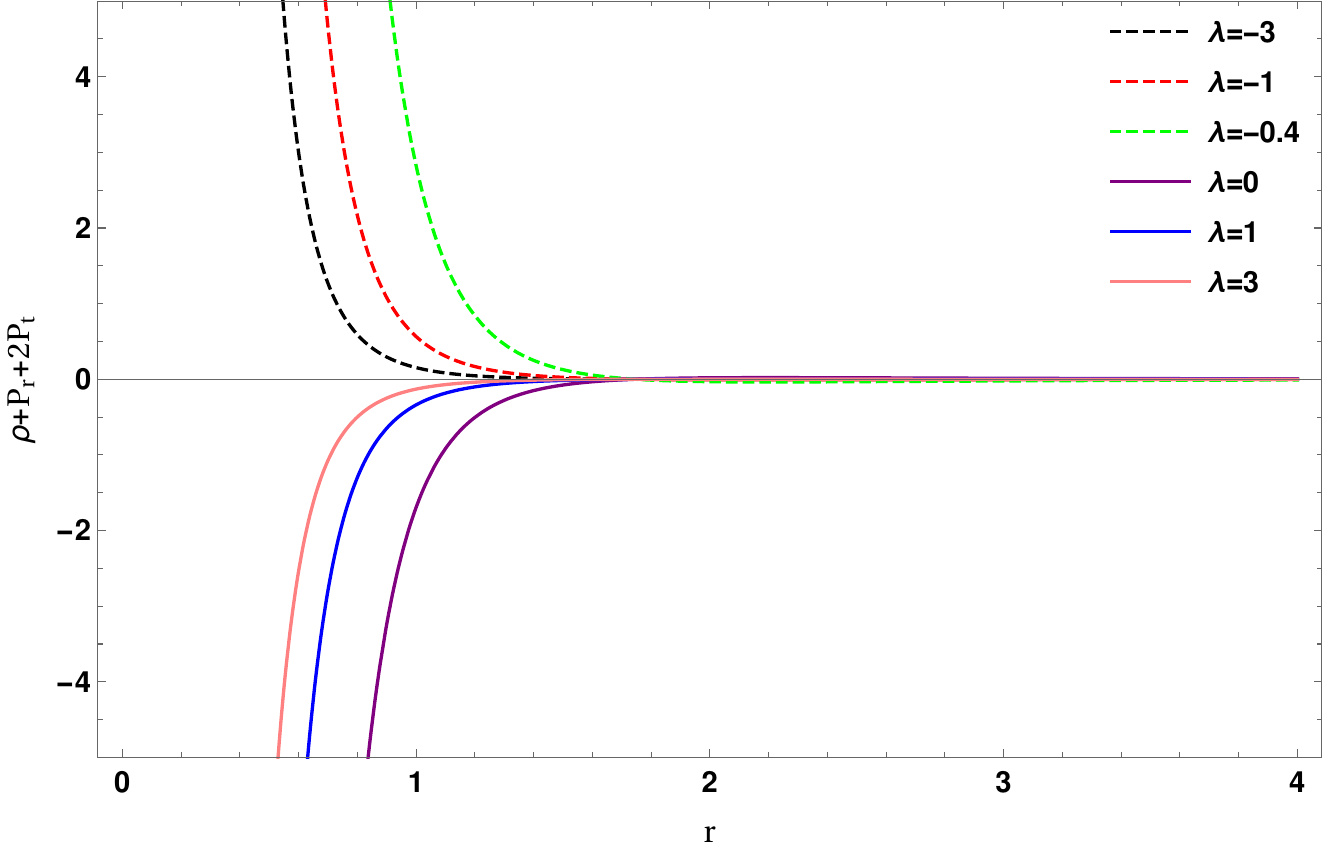}
    \caption{Plots for energy conditions terms with respect to $r$ for different $\lambda$ values}
    \label{energyconditions3}
\end{figure}
Similarly I can calculate the energy conditions terms such as $\rho+|P_r|$, $\rho+|P_t|$, $\rho+P_r+2P_t$ etc using Eq.~(\ref{rho3}-\ref{pt3}).
\subsection{Case 4: $\phi(r)=\phi_0$, $\omega(r)=1+\ln{(r)}$}
I have assumed the matter inside the WH follows the relation
\begin{equation}
    P_r+\omega(r)\rho=0
\end{equation}
where$\omega(r)$ is the EoS parameter varying with radial co-ordinates. Wormholes with varying EoS was first introduced in \cite{Rahaman3}. From Eq.~(\ref{redshift}), I am able to obtain,
\begin{equation}
    \omega(r)=\frac{b}{b'r}
    \label{omega}
\end{equation}
I have considered $\omega(r)=1+\ln{(r)}$ with constant red shift function throughout my calculation. Integrating Eq.~(\ref{omega}) will lead to the form of shape function as,
\begin{equation}
    b(r)=\frac{r_0(\ln{(r)}+1)}{\ln{(r_0)}+1}
\end{equation}
\begin{figure}
    \centering
    \includegraphics[width=8cm]{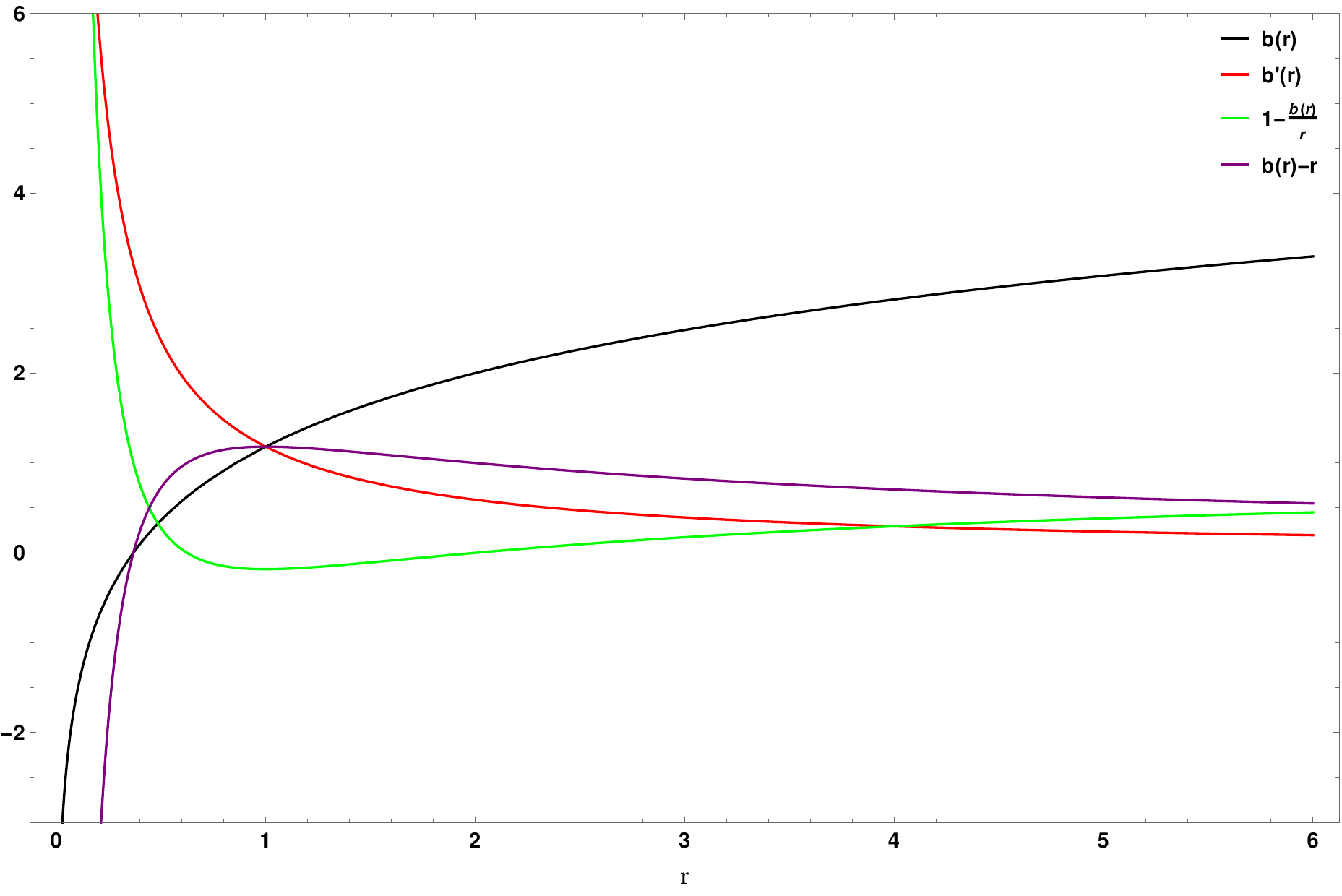}
    \caption{Plot showing the behaviour of shape function $b(r)$, throat condition, flaring out condition and asymptotically flatness condition.}
    \label{b4}
\end{figure}
Using Eq.~(\ref{redshift}) I have obtained
\begin{equation}
\rho=\frac{r_0}{r^3(1+2\lambda)(1+\ln{(r_0)})}, ~~P_r=-\frac{r_0(1+\ln{(r)})}{r^3(1+2\lambda)(1+\ln{(r_0)})}, ~~P_t=\frac{r_0\ln{(r)}}{2r^3(1+2\lambda)(1+\ln{(r_0)})}
\label{rho4}
\end{equation}
NEC for this case reads as,
\begin{equation}
    \rho+P_r=-\frac{r_0\ln{(r)}}{r^3(1+2\lambda)(1+\ln{(r_0)})}, ~~ \rho+P_t=\frac{r_0(2+\ln{(r)})}{2r^3(1+2\lambda)(1+\ln{(r_0)})}
\end{equation}
SEC for this case yields $\rho+P_r+2P_t=0$ and DEC gives,
\begin{equation}
    \rho-P_r=\frac{r_0(2+\ln{(r)})}{r^3(1+2\lambda)(1+\ln{(r_0)})}, ~~ \rho-P_t=-\frac{r_0(-2+\ln{(r)})}{2r^3(1+2\lambda)(1+\ln{(r_0)})}
\end{equation}
\begin{figure}
    \centering
    \includegraphics[width=5cm]{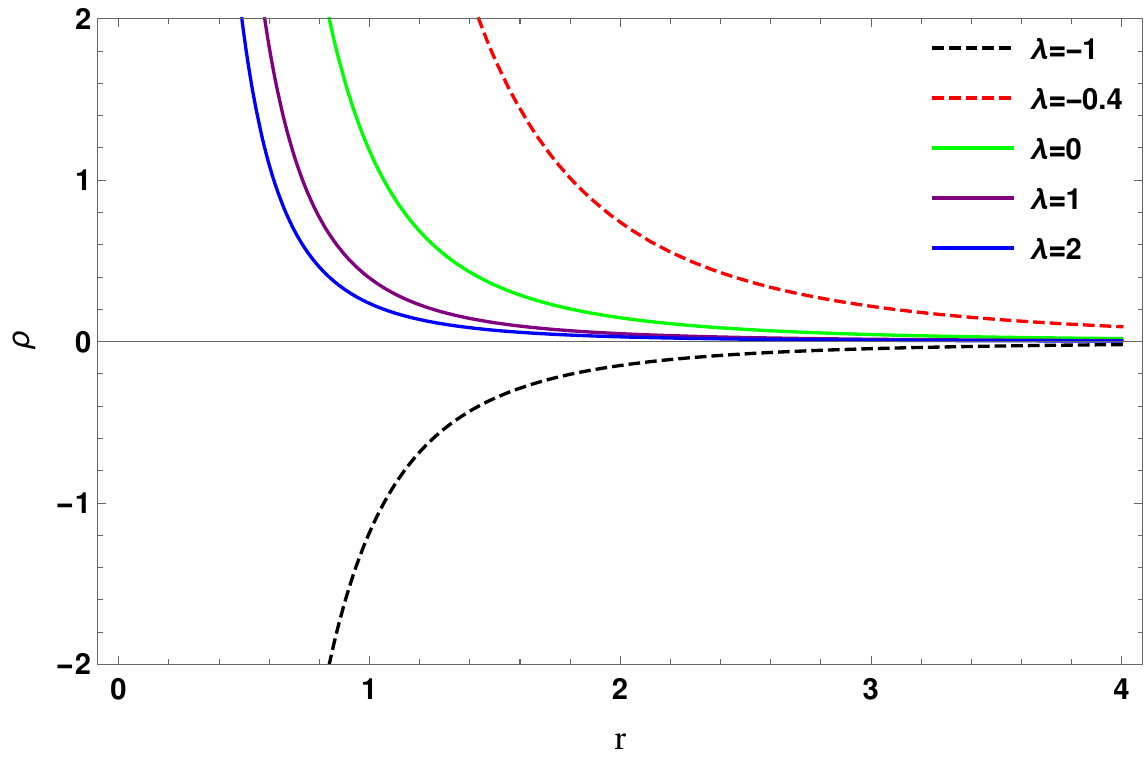}
    \vspace*{0.1in}
    \includegraphics[width=5cm]{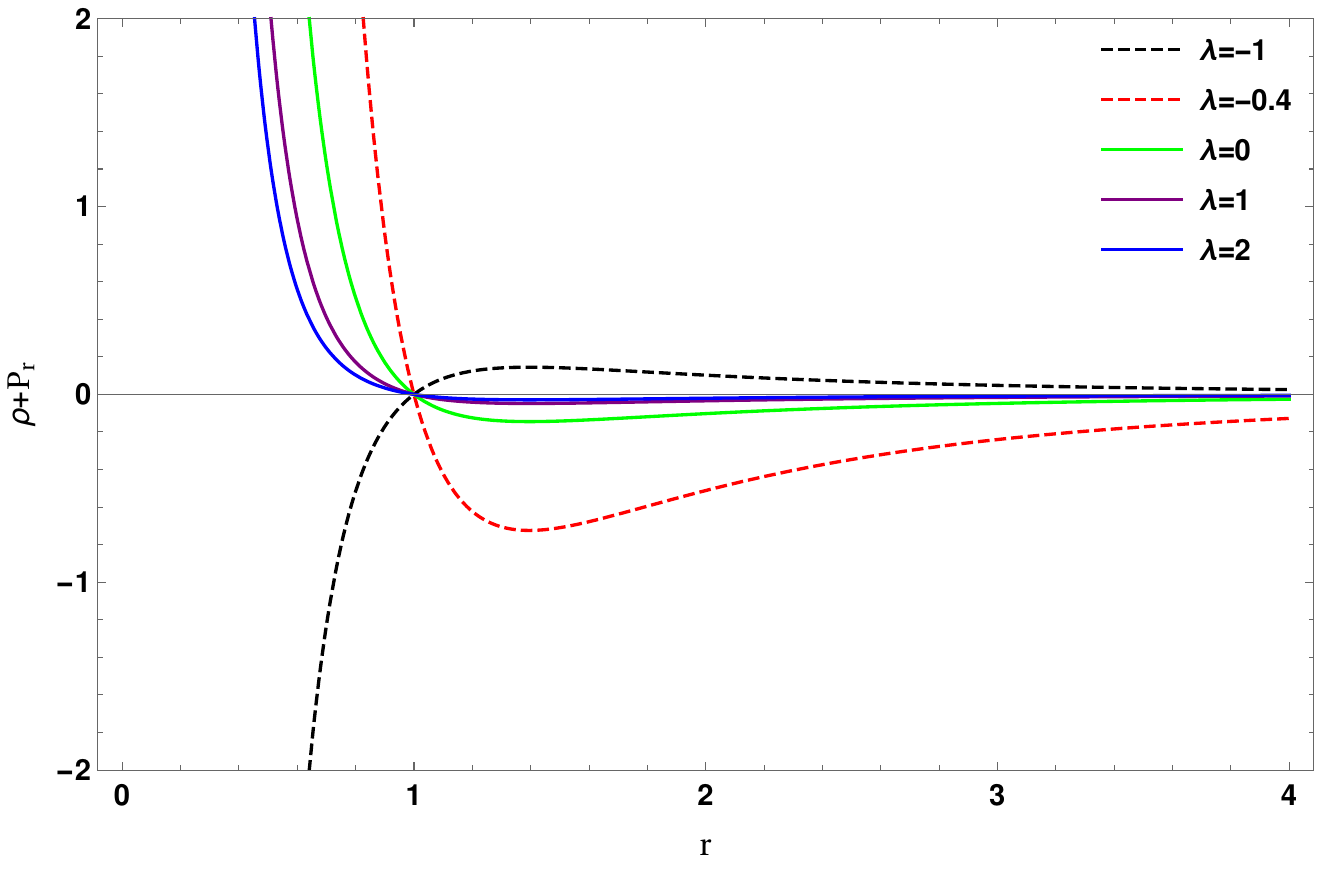}
    \vspace*{0.1in}
    \includegraphics[width=5cm]{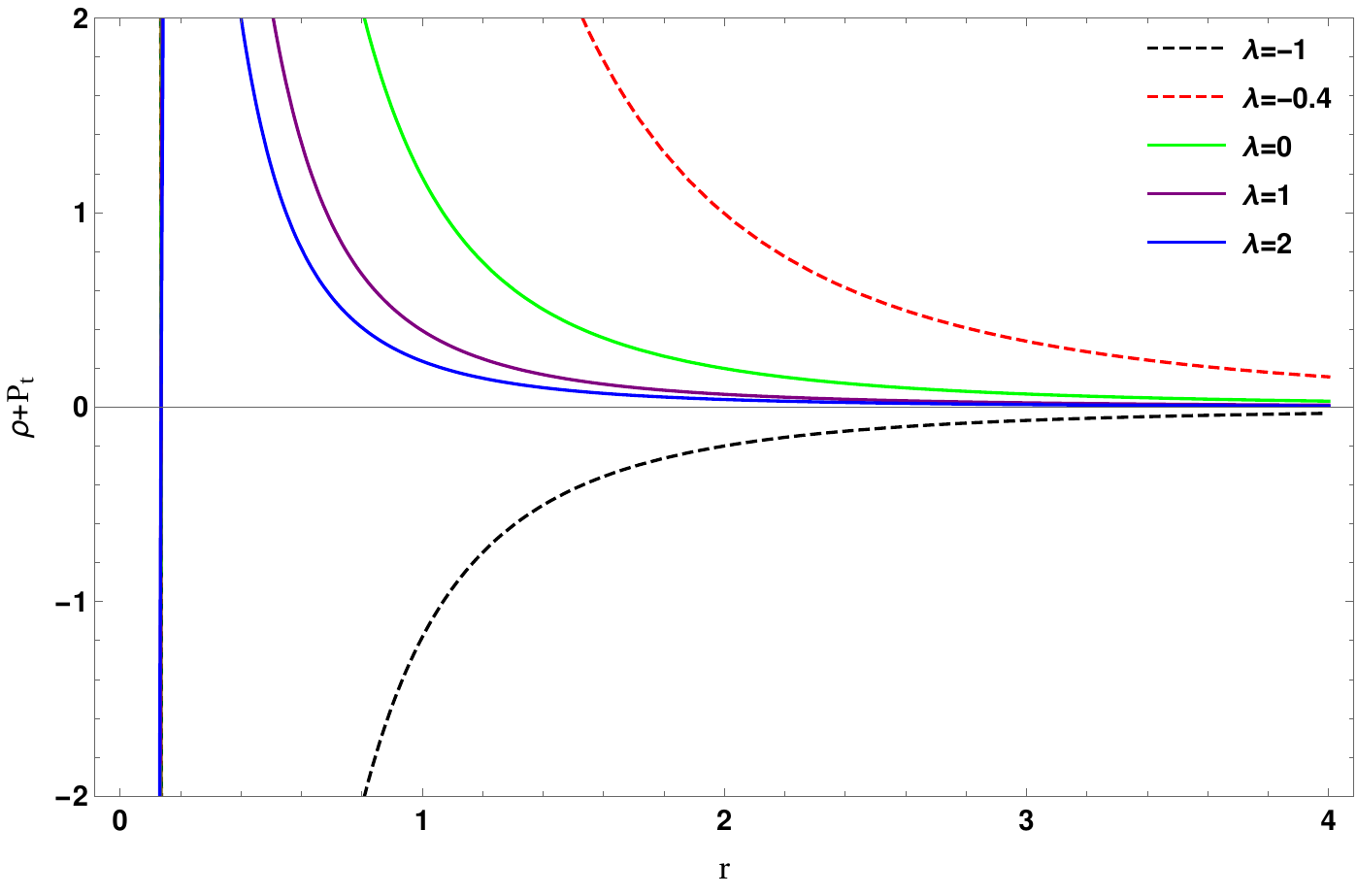}
    \vspace*{0.1in}
    \includegraphics[width=5cm]{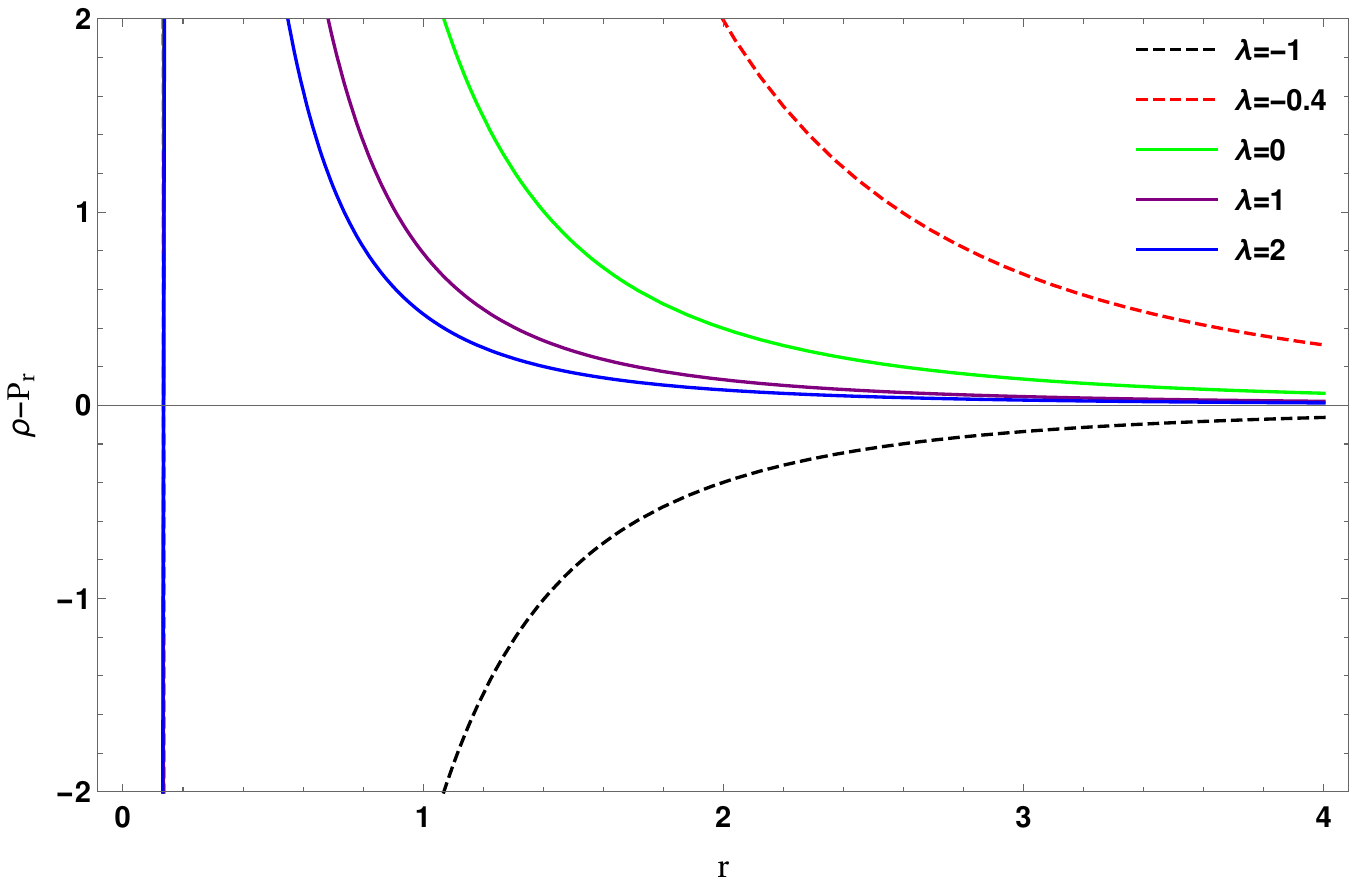}
    \vspace*{0.1in}
    \includegraphics[width=5cm]{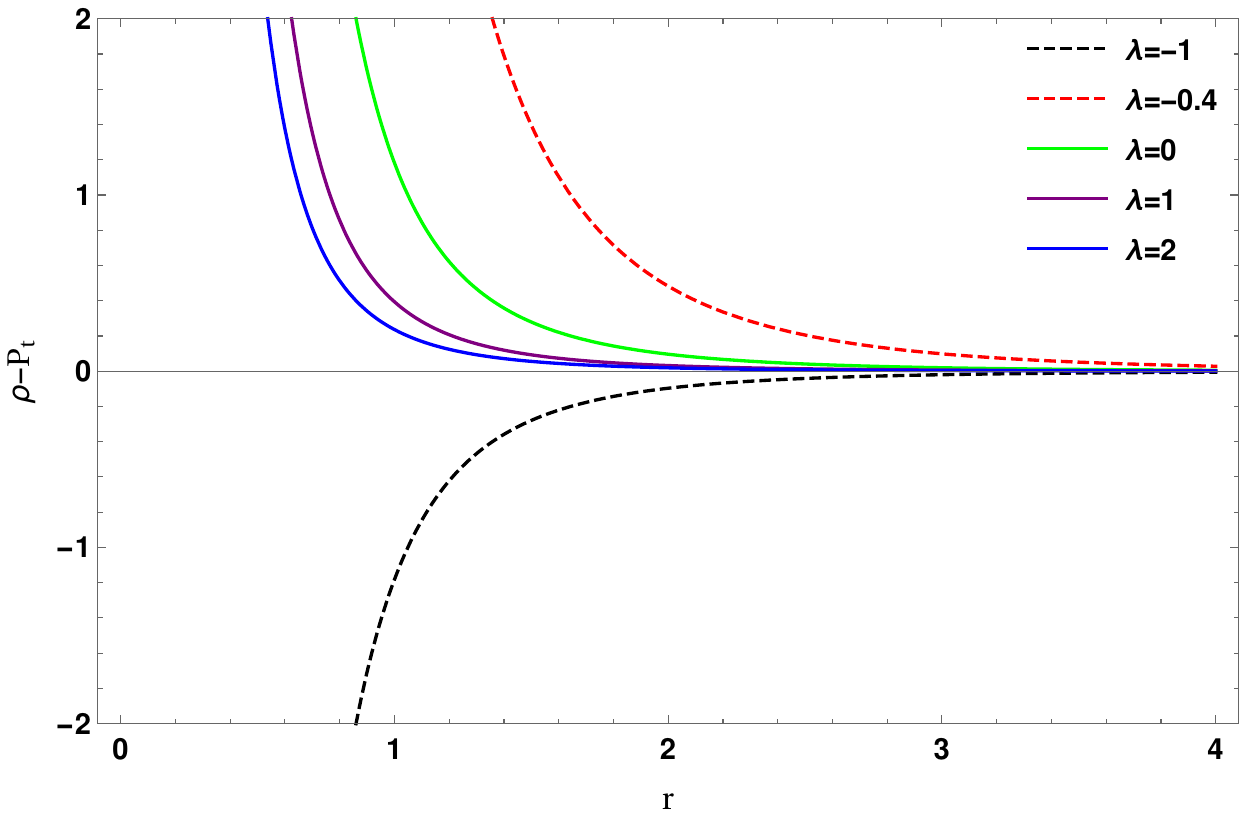}
    \caption{Plots for energy conditions terms with respect to $r$ for different $\lambda$ values}
    \label{energyconditions4}
\end{figure}

\section{Stability Condition:}
 In \cite{capozziello} and \cite{Radhakrishnan} authors have showed the stability analysis of traversable wormhole in modified gravity context. The stability calculation involves the measure of adiabatic sound speed which is expressed as,
\begin{equation}
C_s^2=\frac{\partial P}{\partial\rho}
\end{equation}
Where pressure $P$ is expressed as $P=(P_r+2P_t)/3$.\\ Stability condition requires $$C_s^2=0$$ at the wormhole throat. From Eq.(\ref{rho1}), ~(\ref{pr1}) and (\ref{pt1}), the sound speed for case 1. can be obtained as,
\begin{equation}
\tiny
C_s^2=-\frac{\sech^2{nr}(\phi_0^2(-2 (1 + \lambda) (2 (2 + m n) r + 4 r \cosh{2 n r} -5 m \sinh{2 n r})+4mnr^3(1+4\lambda)(1+nr\tanh{nr})-4mr(1+\lambda)\phi_0(2 n r - (1 - n^2 r^2 + \cosh{2 n r}) \tanh{n r})))}{12(\lambda\phi_0^2(4r+mnr\sech^2{nr}-5m\tanh{nr}+mnr^3(1+4\lambda)\sech^2{nr}(1+nr\tanh{n r})+mr\lambda\sech^2{nr}\phi_0(2nr-(1-n^2r^2+\cosh{2 n r})\tanh{nr})))}
\end{equation}
Similarly for other three models(except case 4), I can calculate $C_s^2$ and the expression are as follows,
\begin{equation}
    C_s^2=\frac{(\lambda +1) \phi _0^2 \left(6 a r_0-5 a r+5 r_0 r-4 r^2\right)-2 a r_0 (4 \lambda +1) r^2+(\lambda +1) r
   \phi _0 (5 a r_0-2 a r+2r_0 r)}{3 \left(\lambda  \phi _0^2 \left(6 a r_0-5 a r+5 r_0 r-4 r^2\right)+2 a r_0 (4
   \lambda +1) r^2+\lambda  r \phi _0 (5 a r_0-2 a r+2 r_0 r)\right)}
\end{equation}
\begin{equation}
\small
    C_s^2=\frac{-r_0 (4 \lambda +1) r^3 a^r \log (a) (r \log (a)-2)+r_0 (\lambda +1) r \phi _0 a^r (r
   \log (a)-2)^2-2 (\lambda +1) \phi _0^2 \left(4 r a^{r_0}-5 r_0 a^r+r_0 r a^r \log
   (a)\right)}{3 \left(r_0 (4 \lambda +1) r^3 a^r \log (a) (r \log (a)-2)+r_0 \lambda  r \phi
   _0 a^r (r \log (a)-2)^2-2 \lambda  \phi _0^2 \left(4 r a^{r_0}-5 r_0 a^r+r_0 r a^r
   \log (a)\right)\right)}
\end{equation}
For case 4, $C_s^2=-1/3$ which suggests that the model is not stable. Considering $C_s^2=0$ at the throat, I have found out that the model is stable for $\lambda=-0.505$ for case 1 where as stability found at $\lambda=-0.314$ and $\lambda=-0.211$ for case 2 and case 3 respectively.

\section{Results and Discussions:}
The concept of Wormhole is still theoretical and have not been observed till now. The $f(R,T)$ theory is helpful in explaining many cosmological problems in different aspects of theoretical physics.The main motivation for working with wormholes within alternative gravity models is to obtain wormhole solutions satisfying the energy conditions, departing from the General
Relativity. There are many literatues available in the context of $f(R,T)$ gravity for wormhole by avoiding the violation of energy conditions.  A wide variety of wormhole solutions has been explored considering different types of shape functions. In this paper, I have considered four different cases to show the behaviour of matter inside the wormhole in $f(R,T)$ theory. In first three cases, I have studied $b(r)= m \tanh{(nr)}, r_0+ar_0\left(\frac{1}{r}-\frac{1}{r_0}\right)$ and $r_0\left(\frac{a^r}{a^{r_0}}\right)$ with $\phi(r)=\frac{\phi_0}{r}$, whereas in last case, I have assumed a new type of EoS parameter varying with radial distance $\omega(r)=1+\ln{(r)}$.  I have shown that all the shape functions follow the throat conditions, flare-out conditions etc.\\
In this section I will now analyze the results related to the presence and the absence of exotic matter in wormhole with the validation of different energy conditions. I have plotted the different energy conditions for various values of $\lambda$ both for positive and negative range.In case 1. I have obtained the expression of different energy conditions parameters as a function of radial distance $r$ and $\lambda$ for $b(r)=m \tanh{(nr)}$ and $\phi=\frac{\phi_0}{r}$. The value of $\lambda$ can not be equal to $-1/4$ and $-1/2$. For the other values of $\lambda$, I have shown the validity of different energy conditions in Fig.~(\ref{energyconditions1}). The energy density is positive for large value of $r$ if $\lambda>-0.25$. The NEC parameter $\rho+P_r$ is negative for $\lambda>-0.5$ whereas $\rho+P_t$ is positive if $\lambda>-0.5$ for  $r\geq2$. The SEC condition $\rho+P_r+2P_t$ is violated for $\lambda>-0.25$ near the throat as well as outside the throat. DEC is violated always for $\lambda>-0.5$. Thus the NEC, SEC, WEC is violated near the throat ($r=2$) if $\lambda>-0.5$ which indicates the presence of exotic matter near the wormhole throat.\\
Further in case 2,  I have observed that energy density is positive for all values of $r$ if $\lambda>-0.5$ from Fig.~(\ref{energyconditions2}). $\rho+P_r$ is negative for $\lambda>-0.5$ and $r>1.5$ whereas $\rho+P_t$ is positive for all values of $r$. On the other hand, $\rho+P_r$ is positive for $\lambda>-0.5$ if $r$ lies between $0.6<r<1.5$. $\rho+P_r+2P_t$ is negative if $\lambda>-0.25$ suggests SEC is violated everywhere. $\rho-P_r$ and $\rho-P_t$ are positive for $\lambda>-0.5$. So NEC, WEC and DEC are violated in all the regions for $\lambda>-0.5$ except  $0.6<r<1.5$. Hence non-exotic matter is present inside the wormhole throat upto $r<1.5$ as NEC, WEC and DEC are satisfied for $0.6<r<1.5$. Violation of all the energy conditions near the throat implies the presence of exotic matter in wormhole.\\

Energy conditions for case 3 is plotted in Fig.~(\ref{energyconditions3}). Energy density is negative for $\lambda>-0.5$ near the throat and $\rho$ goes to zero as $r$ increases beyond the throat radius. $\rho+P_r$ and $\rho+P_t$ are negative for $\lambda>-0.5$ near the throat which demands NEC is violated near the throat. On the other hand SEC is satisfied if $\lambda<-0.25$. Outside the range of $\lambda$, SEC is violated near the throat. DEC is violated near the throat if $\lambda>-0.25$. Lastly in case 4, energy density $\rho$ is positive near the throat $r=2$ if $\lambda>-0.5$ shown in fig.~(\ref{energyconditions4}). Near the throat $\rho+P_r$ is negative for $\lambda>-0.5$ although $\rho+P_r=0$ at $r=1$ and beyond $r<1$, $\rho+P_r$ is positive although $\rho+P_t$ is always positive for $\lambda>-0.5$. So, NEC and WEC is satisfied for $r\leq 1$ if $\lambda>-0.5$. SEC is always satisfied in this particular model as $\rho+P_r+2P_t=0$ everywhere. DEC is also satisfied for $\lambda>-0.5$.\\

 Finally, I can conclude that if $\lambda>-0.25$, all the energy conditions are violated near the throat implies the presence of exotic matter near the WH throat$(r=2)$ for case 1. In Comparison with Rahaman {\it et al} (\cite{Rahaman3}) discussed on $b(r)=m\tanh{(nr)}$ in the context of GR, the presence of exotic matter is confirmed throughout the throat, whereas in $f(R,T)$ gravity we can restrict the existence of exotic matter for different value of $\lambda$.  On the other hand, stability analysis suggests that $\lambda=-0.505$ at the throat and WEC, NEC are satisfied which implies presence of ordinary matter at the wormhole throat at stable point.
 
 For case 2, NEC and WEC is satisfied for $\lambda>-0.5$ within $0.6<r<1.5$. Hence ordinary matter is present within a region inside the throat. However $r\geq 1.5$ all the energy conditions has been violated by the presence of matter which is exotic in nature. Hence the wormhole  shows attractive gravity upto $r>1.5$ and beyond that it is repulsive. Comparing my work with Mishra {\it et al} in (\cite{Mishra3}) for this particular shape function in $f(R)$ gravity context where they have shown a transient behaviour of energy conditions upto $r=1.045$. On the other hand in $f(R,T)$ context the energy conditions are satisfied within $r<1.5$ for a specific range of $\lambda$.  The ordinary matter inside the wormhole throat is also stable as the stable point exist at $\lambda=-0.314$ in this case.
 
 Presence of exotic matter is confirmed in case 3 from the violation of NEC, WEC and DEC if $\lambda>-0.25$ for $f(R,T)=R+2\lambda T$ which is also stable as shown in the stability analysis. The presence of both exotic and ordinary matter has been analyzed for different choices of $f(R)$ in \cite{Mishra3}. Similarly I have shown here the limit of $\lambda$ upto which the matter shows exotic behaviour near the throat in the context of $f(R,T)$. Lastly in case 4, exotic matter is present for $r>1.5$ and $\lambda>-0.5$. But this model is not stable as sound speed is always negative and independent of all the parameters.Thus by assuming different type of shape function or EoS parameter, the wormhole solutions can be obtained with the radius of throat in modified $f(R,T)$ gravity. The exotic nature of matter can be avoided by taking proper values of modified gravity parameter $\lambda$.

\section*{Acknowledgment}
PS would like to thank Department of Science and Technology, Government of India for INSPIRE fellowship. I would also like to thank to Prof. Prasanta Kumar Das and Dr. Gauranga Charan Samanta for useful discussions.


\begin{thebibliography}{10}
\bibitem{Morris} M.~S.~Morris and K.~S. ~Thorne, 
{\it Am. J. Phys.} {\bf 56}, 395 (1988).
\bibitem{Visser}
M.~Visser, 
{\it Springer-Verlag}, New York, 1996).
\bibitem{Visser2} J. Cramer {\it et. al}, {\it Phys Rev. D}, {\bf 51}, 6, (1995).
\bibitem{Lemos} J.~P.~S.~Lemos, F.~S. ~N.~Lobo and S.~Quinet de ~Oliveira, 
{\it Phys. Rev. D } {\bf 68}, 064004 (2003).
\bibitem{Jordan} P.~Jordan, 
{\it Z. Phys.}{\bf 157}, 112 (1959).
\bibitem{Tsukamoto} N.~Tsukamoto, 
{\it Phys. Rev. D} {\bf 95} 8, 084021 (2017)
\bibitem{Tsukamoto2} N.~Tsukamoto, 
{\it Phys. Rev. D} {\bf 94}, 12, 124001 (2016).
\bibitem{Zhou} M.~Zhou, A.~Cardenas-Avendano, C.~Bambi, B.~Kleihaus and J.~Kunz, 
{\it Phys. Rev. D}{\bf 94} 2, 024036 (2016).
\bibitem{Rahaman}  F.~Rahaman, P.~K.~F.~Kuhfittig, S.~Ray and N.~Islam, 
{\it Eur. Phys. J. C} {\bf 74}, 2750 (2014).
\bibitem{Kuhfittig}  P.~K.~F.~Kuhfittig, 
{\it Eur. Phys. J. C} {\bf 74} 99, 2818 (2014).
\bibitem{Bambi}  C.~Bambi, 
{\it Phys. Rev. D} {\bf 87}, 107501 (2013)
\bibitem{Li} Z.~Li and C.~Bambi, 
{\it Phys. Rev. D}{\bf 90}, 024071 (2014).
\bibitem{Nandi} K.~K.~Nandi {\it et al.}, {\it Phys. Rev. D} {\bf 74}024020 (2006)
\bibitem{Harko} T.~Harko {\it et al.}, {\it Phys. Rev. D} {\bf 79}  064001, (2009)
\bibitem{Einstein} A.~Einstein and N.~Rosen, {\it Phys. Rev.} {\bf 48} 73,  (1935).
\bibitem{Hochberg} D. Hochberg and M. Visser, {\it Phys. Rev. D} {\bf56}, 4745,(1997).
\bibitem{Hochberg2}D. Hochberg and M. Visser, {\it Phys. Rev. Lett.} {\bf 81}, 746, (1998).
\bibitem{Morris2} M. S. Morris, K. S. Throne, U. Yurtsever, {\it Phys. Rev. Lett.}, {\bf 61}, 13 (1988).

\bibitem{Jamil}M. Jamil, M. U. Farooq, {\it Int J Theor Phys}, {\bf 49}, 835-841,(2010). \url{https://doi.org/10.1007/s10773-010-0263-z}
\bibitem{Lobo} F. ~S. ~N. ~Lobo, 
{\it Phys. Rev. D} {\bf 71}, 084011, (2005).
\bibitem{Rahaman2} F.~Rahaman,M.~Kalam, M.~Sarker and K. Gayen, {\it Physics Letters B} {\bf 663}, 2 (2006).
\bibitem{Wang}  D. ~Wang D, X.~Meng,{\it Euro. Phys. Jornal C} {\bf 76} 484, (2016). 
\bibitem{Lobo3} F. S. N. Lobo, {\it AIP Conference Proceedings}, {\bf 1458}, 447, (2012).
\bibitem{James} O. James, E. von Tunzelmann, and P. Franklin. K. Throne, {\it American Journal of Physics}, {\bf 83}, 6, (2015), [arXiv:1502.03809]

\bibitem{Rosa} J.~L.~Rosa, J.~P.~S.~Lemos and F.~S.~N.~Lobo, 
{\it Phys. Rev. D} {\bf98} 6, 064054 (2018).
\bibitem{Lobo2} F.~S.~N.~Lobo and M.~A.~Oliveira, 
{\it Phys. Rev. D} {\bf80}, 104012 (2009).
\bibitem{Godani} N. Godani, {\it New Astronomy}, {\bf 94} , 101774,(2022)
\bibitem{Harko2} T.~Harko, F.~S.~N.~Lobo, M.~K.~Mak and S.~V. ~Sushkov, 
{\it Phys. Rev. D} {\bf87}6, 067504 (2013).
\bibitem{sharif} M.~Sharif, Z.~ Zahra,
{\it Astrophys Space Sci} {\bf348}, 275–282 (2013). \url{https://doi.org/10.1007/s10509-013-1545-8}
\bibitem{Zubair} M.~Zubair, S.~Waheed and Y.~Ahmad, 
{\it Eur. Phys. J. C} {\bf76}, 8, 444 (2016).
\bibitem{Godani2} N.~Godani and G.~C.~Samanta, 
{\it Int. J. Mod. Phys. D} {\bf28}, 02, 1950039 (2018).
\bibitem{Godani3} N. ~Godani and G. ~C. ~Samanta, {\it Chinses Journal of Physics},{\bf 62}, 161-171, (2019).
\bibitem{Abbas} G. Abbas, S. Taj, A. Siddiqa, Z. Arbab, {\it Int. J. Geom. Meth. Mod. Phys.} {\bf 20}, 13, 2350236, (2023).
\bibitem{Pradhan} Anirudh Pradhan, Archana Dixit, Akram Ali, Ayan Banerjee, {\it  Int. J. Geom. Meth. Mod. Phys.},{\bf 21}, 12, 2450206, (2024).
\bibitem{Sharif2} M. ~Sharif, A. ~Ikram, {\it Int. J. Mod. Phys. D},{\bf 27}, 01, (2018) [arXiv:1707.05162].
\bibitem{Grezia} E. Di Grezia, E. Battista, M. Manfredonia, and G. Miele, {\it Eur.Phys.J.Plus}, {\bf 132}, 537, (2017), arXiv: 1707.01508.
\bibitem{Battista}E. Battista, S. Capozziello, A. Errehymy, {\it Eur.Phys.J.C}, {\bf 84} (2024) 12, 1314, ArXiv: 2409.09750.
\bibitem{Moreas}P.~H.~R.~S.~Moraes, J.~D.~V.~Arbail and M.~Malheiro, 
{\it JCAP} {\bf 1606}, 005 (2016).
\bibitem{Zubair3} M.~Zubair, G.~Abbas and I.~Noureen, 
{\it Astrophys. Space Sci.} {\bf361}, 1, 8 (2016).
\bibitem{Shamir} M.~F.~Shamir, 
{\it Eur. Phys. J. C} {\bf75}, 8, 354 (2015).
\bibitem{Momeni} D.~Momeni, P.~H.~R.~S.~Moraes and R.~Myrzakulov, 
{\it Astrophys. Space Sci.} {\bf361},7, 228 (2016).
\bibitem{Moraes4} P.~H.~R.~S.~Moraes, G. ~Ribeiro and R.~A.~C.~Correa,
{\it Astrophys. Space Sci.} {\bf361}, 7, 227 (2016).
\bibitem{Moraes5} P.~H.~R.~S.~Moraes and R.~A.~C.~Correa, 
{\it Astrophys. Space Sci.} {\bf 361}, 3, 91 (2016).
\bibitem{Correa} R.~A.~C.~Correa and P.~H.~R.~S.~Moraes, 
{\it Eur. Phys. J. C} {\bf76}, 2, 100 (2016).
\bibitem{Alvarenga} F. G. Alvarenga, A. de la Cruz-Dombriz, M. J. S. Houndjo, M. E. Rodrigues and D. SezGmez, 
{\it Phys. Rev. D} {\bf87}, 10, 103526 (2013)

\bibitem{Myrzakulov} R. Myrzakulov,
{\it Eur. Phys. J. C} {\bf72}, 2203 (2012)

\bibitem{sharma} U. K. Sharma and A. Pradhan, 
{\it Int. J. Geom. Meth. Mod. Phys.} {\bf15}, 01, 1850014 (2017).
\bibitem{Nagpal} R. Nagpal, S. K. J. Pacif, J. K. Singh, K. Bamba and A. Beesham, 
{\it Eur. Phys. J. C} {\bf78}, 11, 946 (2018).

\bibitem{Debnath} P. S. Debnath, 
arXiv-preprint, arXiv:1907.02238 (2019).

\bibitem{Ahmed} N. Ahmed and S. Z. Alamri, 
{\it Res. Astron. Astrophys.} {\bf18}, 10, 123 (2018).



\bibitem{Mishra} A.~K.~Mishra, U.~K.~Sharma, V.~C.~Dubey {\it et al.} 
{\it Astrophys Space Sci} {\bf365}, 34 (2020). \url{https://doi.org/10.1007/s10509-020-3743-5}
\bibitem{Sahoo} P.~K.~Sahoo, P.~H.~R.~S.~Moreas, {\it Int. J. Mod. Phys.} {\bf 28}, 15, (2019).
\bibitem{Sahoo2} P.~Sahoo, A.~Kirschner, P.~K.~Sahoo,{\it Mod. Phys. Letter A},{\bf 34} ,37, (2019).

\bibitem{Parbati} P.~Sahoo, S.~Mandal, P.~K.~Sahoo,{\it New Astronomy}, {\bf 80}, 101421, (2020).
\bibitem{Samanta} G. C. Samanta, N. Godani and K. Bamba, 
arXiv:1811.06834v1 [gr-qc] (2018).
\bibitem{Mishra3} A.~K.~Mishra, U.~K.~Sharma,{\it New Astronomy},{\bf88},101628, (2021).
\bibitem{Parsaei} F. Parsaei and S. Rastgoo, arXiv:2110.07278[gr-qc],(2022).
\bibitem{Mishra4} B.~Mishra, A.~S.~Agarwal, S.~K.~Tripathy, S.~Ray, {\it Int. J Mod. Phys. D}, {\bf 30}, 08, (2021).
\bibitem{Elizalde} E. Elizalde and M. Khurshudyan, 
{\it Phys. Rev. D} {\bf98}, 123525 (2018)
\bibitem{Bhatti1} M.Z. Bhatti, Z. Yousaf, M. Nazeer, {\it Int.J.Geom.Meth.Mod.Phys.}, {\bf 20}, 14, 2450001, (2023).
\bibitem{Bhatti} M.~Z.~Bhatti, Z.~Yousaf and M.~Ilyas, 
{\it J. Astrophys. Astr.} {\bf39} 69 (2018).
\bibitem{Moraes3} P.~H.~R.~S.~Moraes, W.~de ~Paula and R.~A.~C.~Correa, 
{\it Int. J. Mod. Phys. D} {\bf28}, 08, 1950098 (2019).
\bibitem{Garcia} N.~M.~Garcia and F.~S.~N.~Lobo, 
{\it Phys. Rev. D} {\bf82}, 104018 (2010).
\bibitem{Garcia2} N.~M.~Garcia and F.~S.~N.~Lobo, 
{\it Class. Quant. Grav.} {\bf28}, 085018 (2011).
\bibitem{Elizalde2} E. Elizalde and M. Khurshudyan, 
{\it Phys. Rev. D} {\bf99},2, 024051 (2019).

\bibitem{Moraes6} P.~H.~R.~S.~Moraes and P.~K. Sahoo {\it Phys. Rev. D} {\bf96}, 044038, 2017.

\bibitem{Moraes7} P.~H.~R.~S.~Moraes and P.~K. Sahoo,{\it  Eur. Phys. J. C} {\bf79}, 677 (2019). \url{https://doi.org/10.1140/epjc/s10052-019-7206-5}
\bibitem{Rahaman3} F. Rahaman {\it et al.}, {\it Acta Phys. Polon. B}, {\bf40}, 1 (2009).
\bibitem{capozziello} S. Capozziello, O. Luongo, L. Mauro, {\it EPJC}, {\bf 136}, 2, (2012),ArXiv: 2012.13908
\bibitem{Radhakrishnan} R. Radhakrishnan,  P. Brown, J. Matulevich, E. Davis, D. Mirfendereski, and G. Cleaver, {\it symmetry}, {\bf 16}, 8, (2024), arXiv:2405.05476


\end{thebibliography}
\end{document}